\documentclass[aps,reprint,pra,letterpaper,floatfix,nobalancelastpage,notitlepage]{revtex4-1}
\usepackage{amsmath}
\usepackage{amsfonts}
\usepackage{graphicx}
\usepackage{bm}
\usepackage{color}
\usepackage{braket}
\usepackage{bbold}
\usepackage[utf8]{inputenc}

\usepackage{color}

\begin{document}

\title{Continuous and time-discrete non-Markovian system-reservoir interactions: Dissipative coherent quantum feedback in Liouville space}

\author{Oliver Kaestle}
\email{o.kaestle@tu-berlin.de}
\affiliation{Technische Universit\"at Berlin, Institut f\"ur Theoretische Physik, Nichtlineare Optik und Quantenelektronik, Hardenbergstra\ss e 36, 10623 Berlin, Germany}
\author{Regina Finsterhoelzl}
\affiliation{Technische Universit\"at Berlin, Institut f\"ur Theoretische Physik, Nichtlineare Optik und Quantenelektronik, Hardenbergstra\ss e 36, 10623 Berlin, Germany}
\author{Andreas Knorr}
\affiliation{Technische Universit\"at Berlin, Institut f\"ur Theoretische Physik, Nichtlineare Optik und Quantenelektronik, Hardenbergstra\ss e 36, 10623 Berlin, Germany}
\author{Alexander Carmele}
\affiliation{Technische Universit\"at Berlin, Institut f\"ur Theoretische Physik, Nichtlineare Optik und Quantenelektronik, Hardenbergstra\ss e 36, 10623 Berlin, Germany}
\date{\today}

\begin{abstract}
Based on tensor network realizations of path integrals reducing exponential memory scaling to polynomial efficiency and a Liouville space implementation of a time-discrete quantum memory, we investigate a quantum system simultaneously exposed to two structured reservoirs. For this purpose, we employ a numerically exact quasi-2D tensor network combining both diagonal and off-diagonal system-reservoir interactions with a twofold memory for continuous and discrete retardation effects. As a possible example, we study the non-Markovian dynamical interplay between discrete photonic feedback and structured acoustic phonon modes, resulting in emerging inter-reservoir correlations and long-living population trapping within an initially-excited two-level system.
\end{abstract}

\maketitle

In spite of recent advances in the field of non-Markovian system-reservoir interactions and novel developments of involved numerical approaches for open quantum systems, their accurate description remains an immense challenge~\cite{Schroeder2019, Luchnikov2019, Luchnikov2020, Kuhn2019}. To deal with the exponential scaling of the Hilbert space dimension, multiple theoretical perspectives have been fathomed, ranging from second-order perturbative master equations~\cite{Breuer2016, Vega2017} and correlation expansions~\cite{Vagov2011pssb, Glaessl2011} to numerically exact real-time path integral formulations for pure decoherence~\cite{Feynman1965, Weiss2011, Caldeira1983, Vagov2011, Glaessl2013, Barth2016, Cosacchi2018, Makri1995, Makri1995a}.
So far, these approaches have focused on the interactions of a quantum system with a single structured reservoir, either in the form of a continuous bath or a time-discrete memory, resulting in time-delayed information backflow to the system~\cite{Budini2018, Taranto2019, Pollock2018, Pollock2018a, Li2018, Li2019, delPino2018, Regidor2020, DenningBundgaard2020}.


\begin{figure}[b]
\centering
\includegraphics[width=\linewidth]{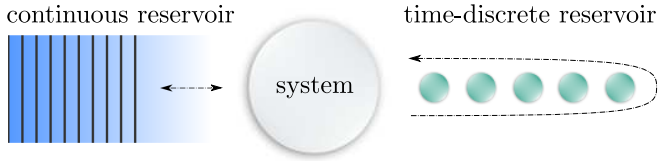}
\caption{Sketch of the considered open quantum system, coupled to both a continuous and a time-discrete structured reservoir.}
\label{fig:system}
\end{figure}

\begin{figure*}[t]
\centering
\includegraphics[width=\textwidth]{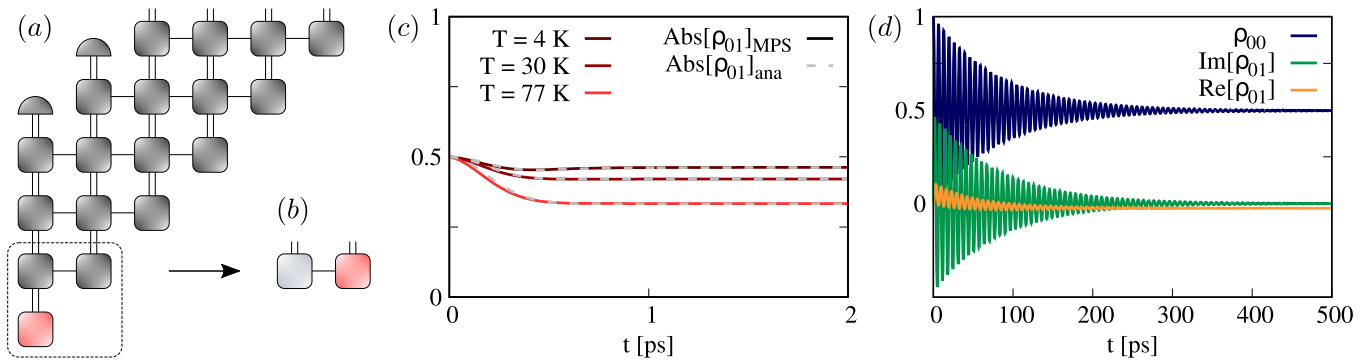}
\caption{(a) Efficient tensor network implementation of real-time path integrals, with the current and past system states stored in individual tensors. (b) System MPS containing current (red) and preceding states (grey) after the first network contraction. (c) Path integral calculations of the \textit{Independent Boson model} at varying temperatures (solid lines) and compared to its analytical solution (dashed lines). (d) Path integral calculation of the \textit{Spin Boson model} dynamics at a reservoir memory depth $n_c=100$ and parameters $\Omega_0= 0.5\,\mathrm{ps}^{-1}$ and $T=77\,$K for the driving field and temperature, respectively.}
\label{fig:mps_harmonic}
\end{figure*}


In this work, we present a numerically exact tensor network-based approach allowing for the first time to describe two non-Markovian processes simultaneously, namely continuous and discrete retardation effects where interactions with both diagonal and off-diagonal system couplings are taken into account, i.e., couplings without or with energy exchange between system and reservoir.
Recently established matrix product state (MPS) techniques to implement a time-discrete memory were aimed at quantum systems where decoherence and dephasing effects are not a key factor, and thus were based on a wave function ansatz to solve the quantum stochastic Schrödinger equation~\cite{Pichler2016, Guimond2016, Guimond2017}. We extend this approach to a time bin-based density matrix description \textit{in Liouville space} to include Markovian and non-Markovian decoherence effects. In a second step, it is combined with a tensor network-based real-time path integral method to describe interactions with a continuous structured reservoir~\cite{Strathearn2017, Strathearn2018, Jorgensen2019, Gribben2020, Strathearn2020thesis}, resulting in a quasi-2D tensor network formalism~\cite{Eisert2019, Tamascelli2019, Finsterhoelzl2020, FinsterhoelzlCarmele2020}.
This architecture enables non-Markovian simulations of quantum systems coupled to two structured reservoirs, see Fig.~\ref{fig:system}, maintaining the relevant entanglement information and capturing both diagonal and off-diagonal system-reservoir interactions on equal footing. Possible applications include setups of waveguide-QED with dephasing~\cite{Calajo2019, Crowder2020, WangRabl2020, Regidor2020}, e.g. realized by additional decay channels, or multiple spatially separated solid-state quantum emitters initially prepared in a dark state and interacting with their environment~\cite{CarmeleNemet2020}.
Here, we specifically consider a two-level quantum system coupled to a structured reservoir of independent oscillators and subject to time-discrete coherent quantum feedback, extending the paradigm of the \textit{Spin-Boson model} to the feedback realm~\cite{Leggett1987}. We demonstrate that non-Markovian interplay between relaxation and decoherence processes results in a dynamical protection against destructive interference and thereby enables population trapping.
This expands upon the widely-discussed localized phase stabilization in the \textit{spin-boson model}~\cite{Leggett1987, Strathearn2018} from an incoherent feedback-induced perspective, replacing the coherent driving with another structured reservoir.

The paper is organized as follows: In Sec.~\ref{sec:continuous}, a recently established MPS-based path integral implementation for continuous reservoirs is discussed. Afterwards, we introduce an MPS implementation of a time-discrete quantum memory in Liouville space in Sec.~\ref{sec:discrete}. In Sec.~\ref{sec:2dnetwork}, we combine the two algorithms to form a quasi-2D tensor network, enabling numerically exact calculations of two non-Markovian system-reservoir interactions, before demonstrating its capabilities in Sec.~\ref{sec:results}, where we find a dynamical protection of coherence in the presence of two non-Markovian reservoirs. Lastly, we summarize our findings in Sec.~\ref{sec:conclusions}.

\section{Path integral formulation for continuous reservoirs} \label{sec:continuous}

We start with the implementation and evolution of a system subjected to a continuous harmonic reservoir. For a numerically exact description, our theoretical approach is based on a real-time path integral formulation~\cite{Feynman1965, Weiss2011, Caldeira1983, Vagov2011, Glaessl2013, Barth2016, Cosacchi2018, Makri1995, Makri1995a}. In recent breakthroughs, path integrals have been implemented in a tensor network approach based on MPS~\cite{Strathearn2018, Jorgensen2019}, allowing to solve non-Markovian dynamics and providing an efficient representation using high-dimensional tensors with restricted correlations~\cite{Strathearn2018}. In the following, we briefly recapitulate the algorithm introduced by \textit{Strathearn et al.}~\cite{Strathearn2017, Strathearn2018, Strathearn2020thesis}, which is employed as a part of our solution to multiple non-Markovian system-reservoir interactions.

Our goal is to employ path integrals for a numerically exact solution of the von-Neumann equation given a Hamiltonian $H(t)$ describing a time-dependent system-reservoir interaction~\cite{Breuer2002, Mukamel1999},
\begin{equation}
\dot{\rho} (t) = \mathcal{L}(t) \rho(t)= -i/\hbar[H(t),\rho(t)],
\label{eq:von_Neumann}
\end{equation}
with $\rho(t)$ denoting the density matrix and $\mathcal{L}(t)$ the Liouvillian superoperator.
As a main challenge, the evaluation of path integrals becomes increasingly expensive over time, since the history of all preceding paths at times $0, \ldots, t_{n-1}$ must be taken into account for the calculation of the current time step $t_n$. However, in case the system-reservoir correlations are finite in time, the \textit{augmented density tensor} scheme can be introduced for improved numerical accessibility~\cite{Makri1995, Makri1995a}.
Exploiting the finite reservoir memory length, only the last $n_c$ time steps are taken into account for the calculation of the current path. This treatment is known as the \textit{finite memory approximation} and results in the augmented density tensor representation as a solution to the system part of Eq.~\eqref{eq:von_Neumann} with traced out reservoir contributions, which reads at time $t_N=N\Delta t$
\begin{align}
\rho_{i_N i^\prime_N} (t_N) &= \prod_{n=1}^N \sum_{i_{n-1}} \sum_{i^\prime_{n-1}} M_{i_n i_{n-1}} M^*_{i^\prime_{n-1} i^\prime_n} \nonumber \\
& \times \prod_{m=n-n_c}^n \exp{ \left( S^{i_n i_m}_{i^\prime_n i^\prime_m} \right) } \rho_{i_0 i^\prime_0} (0),
\label{eq:adt}
\end{align}
and constitutes a discrete path integral formulation where indices $i_n^{(\prime)}$ contain the left (right) configuration of the system at time $t_n=n \Delta t$.
The field transformation matrix $M_{i_n i_{n-1}}$ e.g. accounts for the action of an external driving field $\Omega_0$. For the case $\Omega_0=0$ considered below, it takes the simple form $M_{i_n i_{n-1}}= \mathbb{1}\delta_{i_n,i_{n-1}}$. The influence functional is given by
\begin{equation}
S^{i_n i_m}_{i^\prime_n i^\prime_m} =  -( i_n - i^\prime_n ) [ \eta_{n-m} i_m - \eta^*_{n-m} i_m^\prime ],
\end{equation}
with 
\begin{equation}
\eta_{n-m} = \int_{(n-1)\Delta t}^{n \Delta t} \mathrm{d} \tau \int_{(m-1)\Delta t}^{m \Delta t} \mathrm{d} \tau^\prime \, \phi(\tau-\tau^\prime),
\end{equation}
and $\phi(\tau-\tau^\prime)$ the reservoir autocorrelation function~\cite{Weiss2011, Caldeira1983, Leggett1987}. Using an \textit{improved} finite memory approximation, for $n-m \equiv n_c$ all former paths up to $t_{n_c}=n_c \Delta t$ are additionally incorporated in the integration, i.e. $\eta_{n_c} := \eta_{n-m} + \sum_{k=1}^{n-n_c-1} \eta_{n-k}$~\cite{Strathearn2017}.
Under this approximation, it is possible to restate the augmented density tensor and its time evolution efficiently as a tensor network~\cite{Strathearn2018}: First, Eq.~\eqref{eq:adt} is mapped to a vector $\rho_{j_n}$ in Liouville space,
\begin{align}
\rho_{j_N} (t_N) &= \prod_{n=1}^N \prod_{m=n-n_c}^n I(j_n,j_m) \rho_{j_0} (0),
\end{align}
with $I(j_n,j_m) := \sum_{j_{n-1}} \tilde{M}_{j_n j_{n-1}} \exp{ ( \tilde{S}^{j_n j_m} ) }$. Here, left and right system indices $i_k$, $i_k^\prime$ have been combined to a single index $j_k$ for each time step, resulting in Liouville space representations $\tilde{M}_{j_n j_{n-1}}$ and $\tilde{S}^{j_n j_m}$ of the field transformation matrix and the influence functional, respectively. Afterwards, the augmented density tensor is rewritten as an MPS, storing the present and up to $n_c-1$ past states in individual tensors with the oldest state located at the left end of the MPS. In this representation, tensor compression by consecutive applications of the singular value decomposition~\cite{Schollwoeck2011} reduces the memory requirements to polynomial rather than exponential scaling with respect to $n_c$~\cite{Strathearn2018}.

\begin{figure*}[t]
\centering
\includegraphics[width=\textwidth]{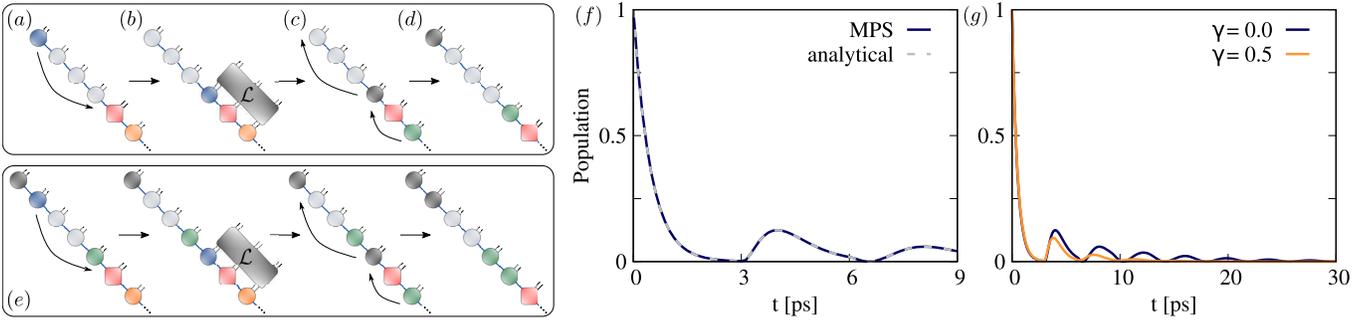}
\caption{MPS implementation of a time-discrete memory in Liouville space. (a)-(d) show the operations performed during the first time step, updating the system state (red square) interacting with present and past reservoir bins (orange and blue circles) by application of the Liouvillian $\mathcal{L}(t)$. (e) shows the second time step with analogous operations. (f) Occupation dynamics of a two-level emitter subjected to feedback [Eq.~\eqref{eq:H_D}], calculated at $\tau=3\,$ps and $\Gamma=31.6\,\mathrm{ps}^{-1}$. Numerical results obtained from the MPS algorithm (solid blue line) are in excellent agreement with the analytical solution for the single excitation case (dashed grey line). (g) MPS feedback dynamics for an additional, analytically not accessible phenomenological dephasing rate $\gamma=0.5\,\mathrm{ps}^{-1}$ (orange line) and without dephasing (blue line).}
\label{fig:mps_feedback}
\end{figure*}

The time evolution is carried out by a network of matrix product operators (MPOs), shown schematically in Fig.~\ref{fig:mps_harmonic}(a) (dark grey shapes). During the first time step, the initial system state $\rho_{j_0} (0)$ (red shape) is contracted with the first MPO in the network [dashed frame in Fig.~\ref{fig:mps_harmonic}(a)]. As a result, the system state is updated and the preceding path is stored to its left, increasing the length of the MPS by one. Fig.~\ref{fig:mps_harmonic}(b) shows the MPS after completion of the first time step. Once step $n=n_c$ is reached, the oldest path in the MPS is summed over by application of a delta tensor [semicircular shape in Fig.~\ref{fig:mps_harmonic}(a)], corresponding to the improved finite memory approximation~\cite{Strathearn2017, Strathearn2018}. At this stage, the MPS length is fixed for the rest of the time evolution. Moreover, for time-independent problems, apart from the index nomenclature the structure of the MPO remains unchanged for all time steps $n \geq n_c$, resulting in an additional performance gain.

To provide an example of a continuous reservoir of non-interacting harmonic oscillators, we consider the Hamiltonian~\cite{Mahan2000}
\begin{align}
H_C/\hbar&= \int \! \mathrm{d}^3 q \, \left[ \omega_q b^\dagger_{\bm{q}} b_{\bm{q}} + g_{\bm{q}} \sigma_{11} \left( b^\dagger_{\bm{q}} e^{i\omega_q t} + \mathrm{H.c.} \right) \right],
\label{eq:H_C}
\end{align}
corresponding to diagonal system coupling without inducing system transitions, with system operators $\sigma_{ij}=\ket{i}\bra{j}$, bosonic annihilation (creation) operators $b_{\bm{q}}^{(\dagger)}$ of reservoir modes with frequency $\omega_q=c_s |\bm{q}|$, $c_s$ the sound velocity, and a mode $\bm{q}$-dependent system-reservoir coupling amplitude $g_{\bm{q}}$. The corresponding correlation function reads
\begin{align}
\phi({\tau-\tau^\prime}) = \int \mathrm{d}^3 q \ g_{\bm{q}}^2 \bigg\{ &\coth \left( \dfrac{\hbar \omega_{\bm{q}}}{2 k_B T} \right) \cos [\omega_q (\tau-\tau^\prime)] \nonumber \\
&- i \sin [\omega_q (\tau-\tau^\prime) ] \bigg\},
\end{align}
with temperature $T$ and $k_B$ the Boltzmann constant.
In the following, we choose a generic coupling element e.g. describing acoustic bulk phonons interacting with a quantum emitter~\footnote{As generic coupling, we choose the acoustic bulk phonon coupling element of GaAs, given by $g_{\bm{q}}^{ii} = \sqrt{ \hbar q/(2 \rho c_s )} D_i \exp [ - \hbar q^2/(4 m_i \omega_i) ]$, resulting in a transition coupling element $g_{\bm{q}} := g_{\bm{q}}^{22} - g_{\bm{q}}^{11}$~\cite{Carmele2019, Foerstner2003, *Foerstner2003pssb, CarmeleMilde2013}. Here, $D_i$ are deformation potentials, $m_i$ denote effective masses, $\hbar \omega_i$ refer to the confinement energies and $\rho$ is the mass density of GaAs, respectively.}.
As a benchmark of the tensor network-based path integral implementation, we first calculate the analytically solvable \textit{Independent Boson model}, consisting of a single two-level emitter subjected to pure dephasing by a structured harmonic reservoir, as described by Eq.~\eqref{eq:H_C}. In Fig.~\ref{fig:mps_harmonic}(c), we prepare the initial polarization at $\rho_{01}(0)=0.5i$ and calculate the resulting dynamics at varying temperatures $T$ (solid lines). The corresponding analytical solution (dashed grey lines) is given by~\cite{Carmele2019}
\begin{align}
\rho_{01}(t) &= \exp \Bigg\{ \int \mathrm{d}^3 q \ \Bigg[ -\dfrac{ig_{\bm{q}}^2}{\omega_q} t + \dfrac{ig_{\bm{q}}^2}{\omega_q^2} \sin (\omega_q t) \nonumber \\
&- \dfrac{g_{\bm{q}}^2}{\omega_q^2} \coth \left( \dfrac{\hbar \omega_q}{2 k_B T} \right) \left[ 1- \cos (\omega_q t) \right] \Bigg] \Bigg\}  \rho_{01} (0),
\end{align}
exhibiting excellent agreement with the numerical results at all considered temperatures. In addition, the employed method features very high performance, enabling reservoir memory depths of $n_c=100$ and beyond. To exemplify the capabilities of the tensor network implementation, we calculate the time evolution dynamics of the \textit{Spin-Boson model}, corresponding to Eq.~\eqref{eq:H_C} with an additional continuous driving field term at amplitude $\Omega_0$, $H_{SBM}= H_C + \Omega_0 (\sigma_{01}+\sigma_{10})$.  Fig.~\ref{fig:mps_harmonic}(d) shows the resulting dynamics at parameters $n_c=100$, $\Omega_0= 0.5\,\mathrm{ps}^{-1}$ and $T=77\,$K for the memory, driving field and temperature, respectively.

\section{Time-discrete memory in Liouville space} \label{sec:discrete}

As a second non-Markovian reservoir, we consider a discrete time-bin based quantum memory. Recently established implementations rely on an MPS-based wave function ansatz to compute the quantum stochastic Schrödinger equation~\cite{Pichler2016, Guimond2016, Guimond2017}. However, this formulation is inherently incompatible with the previously introduced path integral formulation. As a solution to this problem, we present an MPS implementation of a time-discrete quantum memory \textit{in Liouville space}.
Here, the dynamics of the system density matrix is prescribed by a Liouvillian superoperator [see Eq.~\eqref{eq:von_Neumann}], with a Hamiltonian $H_D$ containing the time-delayed system-reservoir coupling, such that interactions occurring at time $t$ couple back into the system and affect its state at a subsequent time $t+\tau$, with $\tau$ the retardation time.
Such a time-discrete coupling e.g. arises in a two-level emitter with states $\ket{0}$, $\ket{1}$ at an energy difference $\hbar \omega_0$, placed in front of a mirror with a round trip time $\tau$. The corresponding Hamiltonian reads
\begin{align}
&H_D / \hbar = \omega_0 \sigma_{11} \nonumber \\
&+ \sqrt{\dfrac{2\Gamma}{\pi}} \int \! \mathrm{d} k \, \sin \left( \dfrac{\omega_k \tau}{2} \right) \left[ \sigma_{10} r_{k} e^{i(\omega_0-\omega_k)t} + \mathrm{H.c.} \right],
\label{eq:H_D}
\end{align}
describing off-diagonal system coupling leading to energy exchange between system and reservoir and system phase relaxation, with system operators $\sigma_{ij}=\ket{i}\bra{j}$, bosonic annihilation (creation) operators $r_k^{(\dagger)}$ of photon modes with frequency $\omega_k=ck$, $c$ the speed of light, and a constant electron-photon coupling amplitude $\Gamma$.
For an efficient evaluation, the dynamics imposed by $\mathcal{L}(t)$ is translated in a time bin-based MPS formalism~\cite{Pichler2016, Guimond2016, Guimond2017, Finsterhoelzl2020} which maintains the relevant system-environment correlations, scaling with $\tau$. In case of additional phenomenological dissipative channels, the Liouvillian can be extended by the standard Lindblad operator~\cite{Breuer2002, Mukamel1999}.

For the MPS implementation of the time-ordered Liouvillian, we start from the formal solution of the system part of Eq.~\eqref{eq:von_Neumann} for the density matrix,
\begin{equation}
\rho(t) = T \exp \left[ \int_0^t \mathrm{d} t^\prime \ \mathcal{L}(t^\prime)  \right] \rho(0),
\label{eq:rho_evo1}
\end{equation}
with $T$ the time-ordering operator. For an MPS-based approach and in analogy to the time-discrete path integral formulation, we restate Eq.~\eqref{eq:rho_evo1} in a time-discrete basis, which reads at time $t_N=N \Delta t$ 
\begin{equation}
\rho(t_N) = L(N,N-1)L(N-1,N-2)\ldots L(1,0),
\end{equation}
with at time discretization $\Delta t$ and with time-bin normalized operators
\begin{equation}
L(n,n-1) = \exp \left[\sqrt{\Delta t} \int_{(n-1)\Delta t}^{n \Delta t} \mathrm{d} t^\prime \ \mathcal{L}(t^\prime)  \right].
\end{equation}
For the MPS evolution of the density matrix during each time step $n$, the discrete Liouvillian time step operator $L(n,n-1)$ is approximated as a \textit{tenth order} series expansion, i.e.,
\begin{equation}
L(n,n-1) = \sum_{m=0}^{10} \dfrac{\sqrt{\Delta t}^m}{m!} \left[ \int_{(n-1)\Delta t}^{n \Delta t} \mathrm{d} t^\prime \ \mathcal{L}(t^\prime) \right]^m.
\label{eq:liou_evo}
\end{equation}

Figs.~\ref{fig:mps_feedback}(a)-(e) show the tensor network scheme for the implementation of the time-discrete memory.
The square red tensor in Fig.~\ref{fig:mps_feedback}(a) contains the system density matrix at the initial time $t=0$. To consider a time-discrete memory, here $n_{d}=4$ circular tensors to its \textit{left} store the reservoir state in Liouville space at preceding times, with the oldest state located on the left end of the MPS (blue).
The reservoir states for all future time steps are initialized to the \textit{right} of the system bin, containing full reservoir entanglement e.g. at finite temperature. In the following we assume an initial vacuum state. Therefore, during each time step a new empty reservoir bin (orange) is added to the MPS from the right, representing the \textit{present} reservoir state [see Fig.~\ref{fig:mps_feedback}(a)].
The memory loop realization explained in detail below introduces the retardation time $\tau = n_d \Delta t$ by the number of initial memory bins $n_{d}$.

The first step of the time evolution is carried out as follows: By applications of the singular value decomposition algorithm~\cite{Schollwoeck2011}, the first memory bin (blue) is pushed to the left of the system (red) while maintaining relevant entanglement information in the swapping procedure [see Fig.~\ref{fig:mps_feedback}(a)]. The Liouvillian operator $L(1,0)$ for the first time step is then applied to the system bin, current memory bin and present reservoir bin, as shown in Fig.~\ref{fig:mps_feedback}(b).
Afterwards, the processed memory bin (grey) is swapped back to its original position and stored for the rest of the time evolution. The updated present reservoir bin (green) is pushed to the left, taking the role of a memory bin [see Fig.~\ref{fig:mps_feedback}(c),(d)]. Fig.~\ref{fig:mps_feedback}(d) shows the MPS after completion of the first time step. The second time step is carried out in the same fashion, as shown in Fig.~\ref{fig:mps_feedback}(e).
After completion of $n_{d}$ time steps, all initial memory bins have been processed. At step $n_{d}+1$, the reservoir bin modified during the first time step [green bin in Fig.~\ref{fig:mps_feedback}(d)] becomes the current memory bin, containing information of a previous system state and setting off reservoir-induced memory effects in the system in complete agreement with the time-ordered problem.

As a first benchmark for the presented time-discrete quantum memory in Liouville space, we calculate the system dynamics imposed by Eq.~\eqref{eq:H_D}. Fig.~\ref{fig:mps_feedback}(f) shows the unfolding emitter population dynamics at a feedback time $\tau=3.0\,$ps and $\Gamma=31.6\,\mathrm{ps}^{-1}$ calculated using the MPS implementation (solid blue line) and compared to its analytical solution up to $t=3\tau$ (dashed grey line). The latter is given by $\left\langle \sigma_{11}(t)\right\rangle=|\left\langle \sigma_{10}(t)\right\rangle|^2$ only valid in the single-excitation regime with 
\begin{equation}
\left\langle \sigma_{01}(t)\right\rangle = \sum_{n=0}^\infty \frac{e^{-\Gamma t}}{n!}
\!
\left[ \Gamma e^{(\Gamma-i\omega_0)\tau} (t \! - \! n \tau) \right]^n \Theta(t \! - \! n \tau),
\label{eq:fb_ana}
\end{equation}
as calculated in \cite{Dorner2002,Kabuss2015,Carmele2020} and shows excellent agreement with the numerical result. In this regime, i.e. $\Gamma\tau\gg1$, the delay in the amplitude governs the dynamics and leads to re-excitations at multiples of the round trip time $\tau$. The phase of the amplitude $\phi=\omega_0\tau$, however, loses importance in the first $\tau$-intervals due to a stronger decay of the mixing terms in the absolute square of Eq.~\eqref{eq:fb_ana}. 
The advantage of our implementation of quantum feedback in Liouville space becomes evident if the impact of phase destroying processes is in question. Up until now, this impact has only been investigated for a special case, finding the emergence of an Ornstein-Uhlenbeck process during the first $\tau$-intervals~\cite{Carmele2020}. These results have been obtained via analytical calculations, limiting the investigation to a small number of feedback intervals. Steady-state scenarios, however, are out of reach in this case as the evaluation of the phase-noise kernels must be done analytically. In our method, these limits have been overcome. Due to the here presented Liouville architecture, additional Lindblad-based dissipation can be easily implemented without increased numerical expense.

In Fig.~\ref{fig:mps_feedback}(g), we present the dynamics of a decaying, intially excited two-level emitter under the influence of quantum coherent feedback and additional phenomenological dephasing at rate $\gamma$, realized by adding a Lindblad dissipator to the Liouvillian [Eq.~\eqref{eq:von_Neumann}] and $H=H_D$,~\cite{Breuer2002, Mukamel1999}
\begin{equation}
\mathcal{D}[ \sqrt{\gamma/2} \tilde{\sigma}_{11}] \rho(t) = \dfrac{\gamma}{2} [ 2 \tilde{\sigma}_{11} \rho(t) \tilde{\sigma}_{11} - \{ \rho(t), \tilde{\sigma}_{11} \}],
\label{eq:lindblad}
\end{equation}
with a redefined system operator in full configuration space, $\tilde{\sigma}_{11} = \mathbb{1}_{D} \sigma_{11} \mathbb{1}_{D}$, including the time-discrete reservoir basis $\mathbb{1}_{D} = \int \mathrm{d} k \sum_{n=0}^\infty \ket{\{ n_k\}} \bra{\{ n_k\}}$.
The time trace shows long time calculations of the emitter population, comparing the cases $\gamma=0$ (blue line) and $\gamma=0.5\,\mathrm{ps}^{-1}$ (orange line). Most importantly, we see that the pure dephasing process becomes important only after the feedback signal re-excitates the emitter and the stabilization of the incoming and outgoing phase comes into play. In the presence of an additional pure dephasing $\gamma \neq 0$, the initial decay process is unchanged but the re-excitation becomes less efficient until only incoherent re-excitation takes place, leading to a faster decay to zero without population trapping, regardless of the choice of $\phi$.

This important result sheds light on the robustness of quantum feedback processes in the presence of additional Markovian dissipation channels. As expected, additional Markovian decoherence leads to a washing out of the signal since a loss of quantum feedback-induced coherence is inevitable. However, this does not have to be the case in the presence of an additional non-Markovian dissipation channel, which we discuss in the following.


\section{Quasi-2D tensor network} \label{sec:2dnetwork}

As a next step, we expand upon the MPS architecture for time-discrete quantum memory in Liouville space by combining it with the previously discussed tensor network-based path integral implementation for continuous harmonic reservoirs, resulting in a \textit{quasi-2D tensor network}.
The technical connection of the networks via link indices which store arising entanglement information enables the numerically exact description of correlation buildup \textit{in between the reservoirs}. When considering a scenario involving two non-Markovian reservoirs not isolated from each other, such inter-reservoir correlations may have fundamental impact on the system dynamics, therefore prohibiting a strict truncation of the arising inter-reservoir entanglement, e.g. in the form of a low Schmidt value cutoff precision $d_{cut}$. As a result of not only two system-reservoir interactions but additionally arising reservoir-reservoir entanglement, the overall grade of entanglement in the system rises \textit{intensively} with respect to the twofold single reservoir case. On the other hand, in setups where two non-Markovian reservoirs are present but do not crucially interact, e.g. via dynamical decoupling, a much more restrictive truncation is possible without cost of accuracy. For the presented results, we have employed a high Schmidt value cutoff precision $d_{cut}=10^{-12}$, such that no relevant entanglement information is lost during the time evolution. The dynamical interplay between the two reservoirs with the system and with each other poses an immense numerical challenge and strongly limits the accessible memory depths in the here considered system: While the two presented algorithms by themselves enable simulations of a single reservoir with deep memories, their combination is accompanied by limitations due to the arising inter-reservoir entanglement. As a result, the combined number of memory bins in the quasi-2D network is limited to $n_c+n_d < 20$ for our model of choice, as is the case in traditional path integral implementations for a single reservoir~\cite{Caldeira1983, Leggett1987, Vagov2011, Glaessl2013, Barth2016, Cosacchi2018}.
However, we stress once more that this limitation is a natural consequence of the high grade of entanglement in between the two reservoirs and the system. With the presented quasi-2D network architecture, we take first steps to unravel the mostly unexplored field of multiple interacting non-Markovian reservoirs by explicitly considering memory-enabled information backflow in between them.

\begin{figure}[t]
\centering
\includegraphics[width=\linewidth]{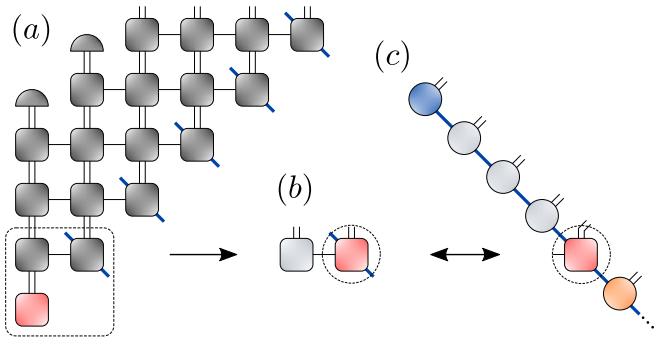}
\caption{Quasi-2D tensor network implementation. (a) Tensor network-based path integral implementation for continuous reservoirs. (b) System MPS containing current (red) and preceding states (grey) after the first network contraction. To realize the quasi-2D network, the current state tensor now features a link connection to the discrete memory MPS (blue diagonal line). (c) Time-discrete memory MPS at the beginning of the first time step, with the common current system state acting as a junction between the reservoirs (dashed circles).}
\label{fig:mps_connect}
\end{figure}

The construction of the quasi-2D network is sketched in Fig.~\ref{fig:mps_connect}: The two tensor networks for the continuous [Figs.~\ref{fig:mps_connect}(a),(b)] and time-discrete reservoirs [Fig.~\ref{fig:mps_connect}(c)] are connected to each other via the common tensor representing the current system state in both MPS algorithms (red shape). The system state tensor is employed to act as a junction connecting the two reservoirs [dashed circles in Figs.~\ref{fig:mps_connect}(b),(c)] and thereby enables the buildup and storage of inter-reservoir correlations in the connecting link indices.
The time evolution of the quasi-2D network is carried out as follows: During each time step, the system is first evolved under the influence of the continuous reservoir by a single contraction of the network, as shown in Fig.~\ref{fig:mps_connect}(a) (dashed frame). Afterwards, the new current system state [red shape in Fig.~\ref{fig:mps_connect}(b)] is subjected to the second tensor network algorithm accounting for the time-discrete reservoir [Fig.~\ref{fig:mps_connect}(c)]. In consequence, the quasi-2D network stores the history of both interactions, maintaining crucial entanglement information and enabling the calculation of two dynamically interacting time-delayed processes.
As an example application for the quasi-2D network, in the following we investigate the interplay of off-diagonal coherent quantum feedback and a diagonal reservoir of independent oscillators. As illustrated below, for certain memory depths and initial states, this results in dynamical protection of coherent quantum feedback properties in the open system.


\section{Memory-induced dynamical population trapping} \label{sec:results}

Coherent quantum feedback mechanisms exhibit a rich variety of non-Markovian phenomena~\cite{Grimsmo2015, Wilson2003, Hetet2011, Hoi2015, Kabuss2015, Svidzinsky2018, Barkemeyer2019}, e.g. enabling coherent population trapping~\cite{Dubin2007, Glaetzle2010, Carmele2013, Guimond2016, Nemet2019, CarmeleNemet2020}, Ornstein-Uhlenbeck-type events in the presence of white noise~\cite{Carmele2020}, and formation of large entangled photon states~\cite{Pichler2017}. However, so far these effects have not been explored in the presence of additional non-Markovian decoherence or dissipative channels.
To investigate the impact of dephasing on feedback-induced decoherence, we consider a two-level emitter placed in front of a mirror with a round trip time $\tau$, taking the role of a time-discrete reservoir (see Fig.~\ref{fig:system}) described by Eq.~\eqref{eq:H_D}.
The photon-induced feedback imprints a time-delayed coherence in the form of a feedback phase $\varphi=\omega_0 \tau / (2 \pi)$ on the system, critically influencing its dynamics. It is given by the delay time $\tau$ and the transition frequency $\omega_0$ of the electronic coherence operator $\sigma_{12}$.
To study a pronounced quantum optical effect, we consider the case of an initially excited two-level system where coherent population trapping occurs as a result of a bound state in continuum at feedback phases $\varphi \in \mathbb{Z}$~\cite{Dubin2007, Glaetzle2010, Carmele2013, Guimond2016, Nemet2019, CarmeleNemet2020, Carmele2020}. The unfolding dynamics are evaluated for parameters $\Gamma=0.9\, \mathrm{ps}^{-1}$, $\tau = 1.2\,$ps and $n_{d}=4$, resulting in a time discretization $\Delta t = 0.3\,$ps, e.g. typical for semiconductor quantum dot based devices~\cite{Carmele2013, Nemet2019, Barkemeyer2019}. Moreover, we get $\Gamma \tau \approx 1.1$, corresponding to the strong non-Markovian regime~\cite{baranger_2019, Pichler2016, Carmele2020}. The employed series expansion of the Liouvillian up to tenth order [see Eq.~\eqref{eq:liou_evo}] justifies this coarse time discretization, making $n_d=4$ time bins sufficient for our investigation while resulting in convergent results (see Appendix~\ref{app:convergence}).

\begin{figure}[t]
\centering
\includegraphics[width=\linewidth]{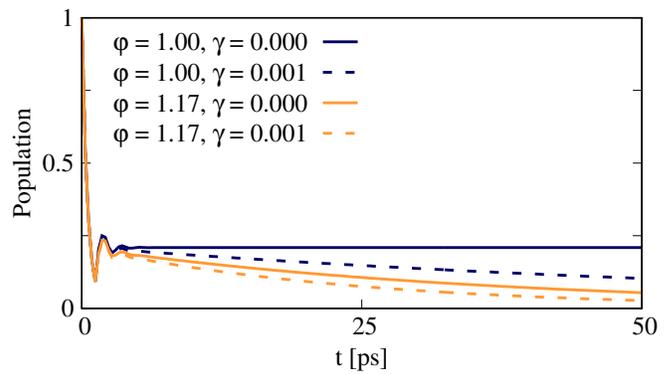}
\caption{Excited state population dynamics of the two-level emitter under time-discrete photon feedback at varying feedback phases $\varphi=\omega_0 \tau/(2\pi)$ and constant dephasing rates $\gamma$.}
\label{fig:lindblad}
\end{figure}

To illustrate the power of our method, we compare the cases of Markovian and non-Markovian dephasing introduced by an additional diagonal system-reservoir coupling, representing the continuous reservoir in Fig.~\ref{fig:system}. As a first step, we calculate the system dynamics given by Eq.~\eqref{eq:H_D} in the presence of phenomenological dephasing at rate $\gamma$, introduced by the Lindblad dissipator stated in Eq.~\eqref{eq:lindblad}.
Fig.~\ref{fig:lindblad} shows resulting excited state population dynamics at varying $\gamma$ and feedback phases $\varphi$. At $\gamma=0$ there exists a periodic ideal feedback phase $\varphi \in \mathbb{Z}$ such that the system decouples from its environment by constructive interference, resulting in coherent population trapping (solid blue line in Fig.~\ref{fig:lindblad}).
Choosing a nonzero dephasing $\gamma=0.001\,\mathrm{ps}^{-1}$ has no impact on the population dynamics until feedback sets in, since the radiative decay is frequency-independent until $t=\tau$ (dashed blue line). Thereafter, phenomenological dephasing destroys the phase interference and with it the trapping mechanism, resulting in an asymptotic decline of occupation to zero.
At a nonideal feedback phase, here $\varphi=1.17$, and no dephasing, destructive interference leads to an asymptotic decline towards zero as well (solid orange line). The decay is further accelerated by setting $\gamma > 0$ (dashed orange line), since any phenomenological decoherence attacking the phase relation $\varphi$ results in faster decay. In conclusion, choosing a feedback phase $\varphi \notin \mathbb{Z}$ without a structured phonon reservoir inevitably results in asymptotic population decay via thermalization, and a Lindblad formulation of decoherence never preserves quantum correlations between the reservoir and system states.


\begin{figure}[t]
\centering
\includegraphics[width=\linewidth]{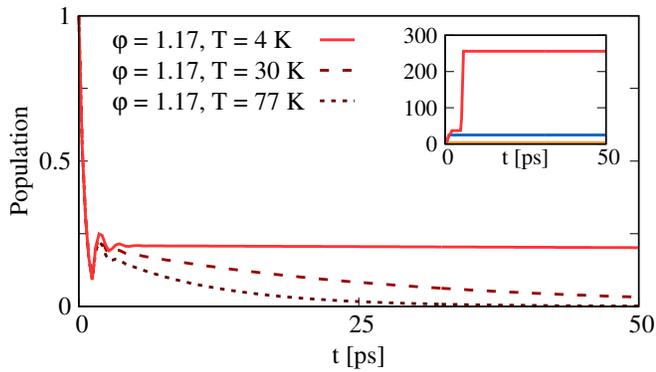}
\caption{Excited state population dynamics of the two-level emitter subject to photon and phonon reservoir interactions at $\varphi \neq 1$ and varying temperatures $T$. Inset: Dimension of the link index connecting the discrete memory MPS to the system in the presence of phonons (red), without phonon coupling (blue) and for phenomenological dephasing (orange).}
\label{fig:path_dynamics}
\end{figure}

As a next step, we show this is not necessarily the case if the decoherence process itself is the result of a non-Markovian reservoir interaction. Using the quasi-2D tensor network, we calculate the emitter dynamics imposed by $H_D+H_C$ [Eqs.~\eqref{eq:H_C},~\eqref{eq:H_D}] at $n_c=4$ and leaving all remaining parameters unchanged. Fig.~\ref{fig:path_dynamics} shows resulting population dynamics at different temperatures. For the chosen parameters, we find population trapping for the non-ideal feedback phase $\varphi=1.17$, i.e. $\varphi \notin \mathbb{Z}$, at $T=4\,$K (solid line).
Time-delayed excitation backflow from the continuous reservoir of oscillators to the system enables a decoupling from destructive interference with the time-discrete photon environment. This information backflow results in correlation buildup and information exchange between the reservoirs, dynamically protecting feedback-induced coherence in the system for long times.
As long as diffusion processes at finite temperature take place on a comparable time scale as the coherent feedback dynamics, we always find dynamical population trapping by tuning of $\varphi$ after a typical excitation backflow time. At higher temperatures, it is to be expected that thermal properties of the phonon reservoir start to dominate the dynamics: Dashed and dotted lines in Fig.~\ref{fig:path_dynamics} show corresponding thermalization dynamics at $T=30\,$K and $T=77\,$K, respectively, exhibiting population decay. The temperature dependence clearly shows that correlation lengths within the full system-reservoir dynamics are of importance, and the observed effect allows to probe these otherwise inaccessible microscopic environmental properties.
This formation of self-stabilizing dissipative structures is closely related to a localized phase stabilization in the coherent driving case $\Omega_0\neq0$ at Ohmic spectral densities and without photons~\cite{Leggett1987, Strathearn2018}. There, above a critical coupling strength the system transitions into a localized phase with nonzero steady state population rather than decaying to zero. In our case, a similiar phenomenon is established with incoherent feedback instead of coherent external driving, addressing the localized phase stabilization process from a dissipative non-Markovian side.
The complexity of this phenomenon is illustrated in the inset in Fig.~\ref{fig:path_dynamics}, showing the dimension of the link index connecting the current open system bin to the discrete memory MPS over time (see Fig.~\ref{fig:mps_connect}). The red line shows the case including the continuous reservoir. After slowly increasing during times $t < \tau$, it exhibits a vast increase once feedback sets in and quickly reaches a maximum due to finite memory. The high grade of entanglement between the two reservoirs even at arguably low finite memory sizes $n_d=4$, $n_c=4$ underlines the crucial role of the interplay between the two non-Markovian processes for the observed protection of coherence.
Switching off the phonon coupling, $g_{\bm{q}}=0$, results in a much lower maximum link dimension (blue line), as no entanglement between the reservoirs arises. For phenomenological dephasing (orange line), the link connecting system and memory bins has an even lower dimension due to the highly decreased complexity of the then 1D network.


\section{Conclusions} \label{sec:conclusions}

We have presented an MPS algorithm for the description of a time-discrete quantum memory in Liouville space. By combining this technique with a path integral tensor network implementation for continuous non-interacting harmonic reservoirs, we have established a quasi-2D tensor network, allowing for simulations of quantum systems subject to two non-Markovian environments while maintaining crucial entanglement information in the coupled system with both diagonal and off-diagonal system-reservoir interactions.
Due to arising reservoir-reservoir correlations, system correlations scale intensively with respect to the twofold single reservoir case. In consequence, the achievable memory depth is limited to $n_c+n_d < 20$ in our study. However, appropriate tuning of the relevant system and reservoir time scales via the employed parameters still opens up a wide array of accessible systems and scenarios where numerical convergence can be achieved. The next step will be to trace out the time-discrete feedback bins as well after their interaction with the systems' degrees of freedom to further improve numerical efficiency. This will allow for longer delay times and therefore increased time discretizations without changing the qualitative results. Hence, the presented quasi-2D tensor architecture is a first step towards unraveling the mostly unexplored field of multiple interacting non-Markovian reservoirs in a numerically exact fashion.

As an example application, we have demonstrated that the interplay of a structured phonon reservoir and photon feedback can dynamically protect the system from destructive interference by time-delayed backflow of coherence, resulting in dynamical population trapping. Tuning the non-Markovian interactions with respect to each other allows for the formation of inter-reservoir correlations, dynamically preservering feedback-induced coherence in the system.
These findings have implications for the fields of quantum thermodynamics and nonequilibrium physics, as well as dynamical quantum phase transitions with ergodic, entropic or negentropic information exchange, where taking account of such dissipative structures may unravel new phenomena. Future works will aim to advance our architecture to a full 2D representation via projected entangled pair states with combined memory bins, potentially allowing for simulations of multi-level systems at improved time resolutions.


\appendix

\begin{figure}[t]
\centering
\includegraphics[width=\linewidth]{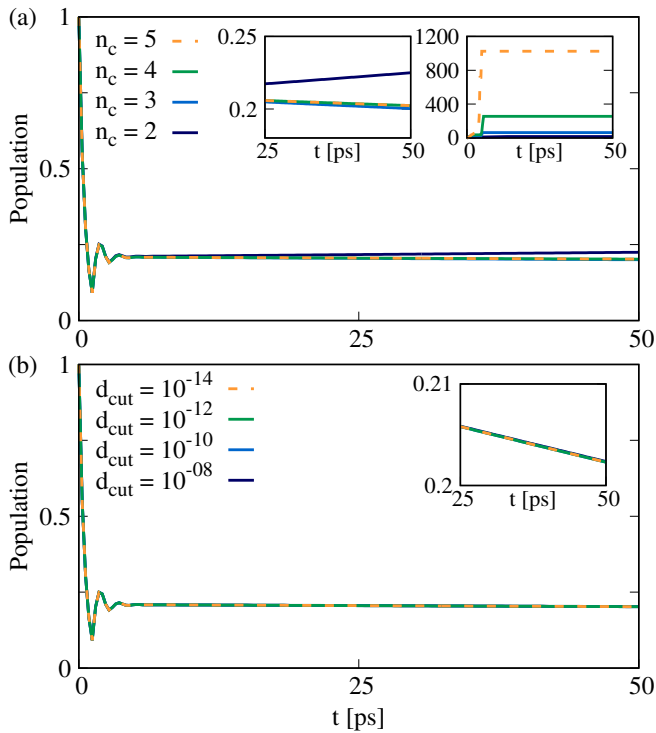}
\caption{Convergence analysis of the results presented in Fig.~\ref{fig:path_dynamics} with respect to (a) the continuous reservoir memory depth $n_c$ and (b) the Schmidt value cutoff precision $d_{cut}$ applied during the singular value decomposition. In (a), the left inset shows a zoom-in on the long term dynamics, with the right inset depicting the resulting dimension of the link index connecting the two tensor networks. The inset in (b) shows a zoom-in on the long term dynamics.}
\label{fig:convergence}
\end{figure}

\begin{figure}[t]
\centering
\includegraphics[width=\linewidth]{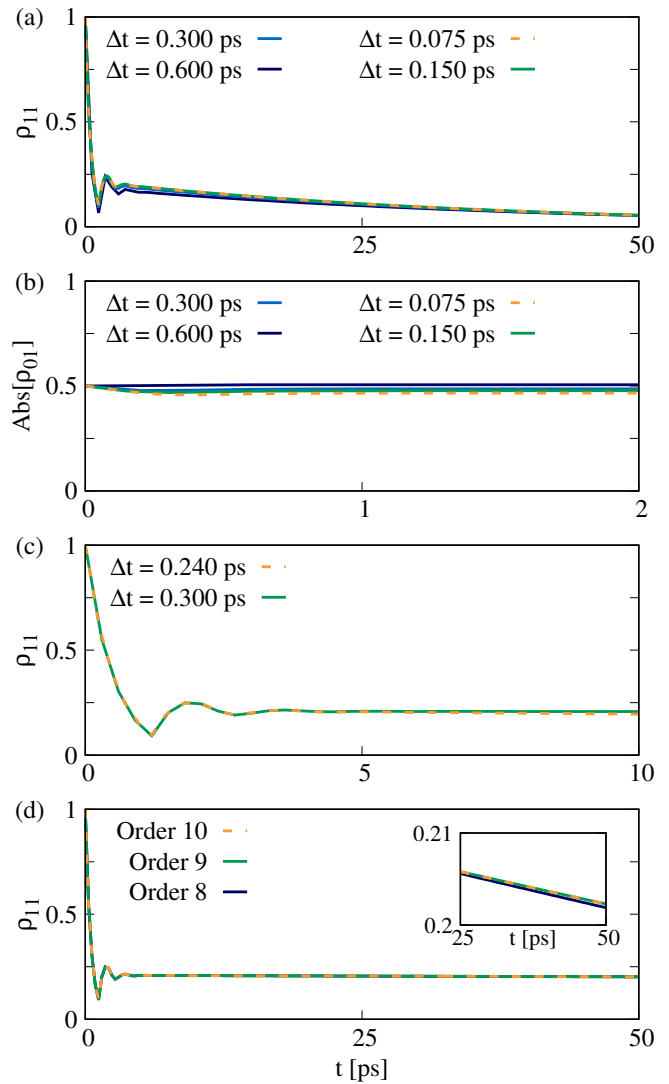}
\caption{Convergence analysis of the presented results with respect to the time evolution step size $\Delta t$ employed in (a) the time-discrete reservoir algorithm (see Fig.~\ref{fig:lindblad}), (b) the time evolution of the continuous reservoir (see Fig.~\ref{fig:mps_harmonic}) and (c) the combined scenario (see Fig.~\ref{fig:path_dynamics}). (d) shows a convergence analysis with respect to the order of the series expansion performed in the Liouvillian $\mathcal{L}(t)$, with the inset showing a zoom-in on the long term dynamics.}
\label{fig:stepsize}
\end{figure}

\section{Convergence analysis} \label{app:convergence}

Here we provide a detailed analysis on the numerical convergence of the presented results. In light of the memory limitations imposed by an intensive scaling of entanglement when describing two non-Markovian reservoirs simultaneously, we first confirm convergence with respect to the memory depth of the continuous harmonic reservoir $n_c$, i.e., the validity of the finite memory approximation for the considered scenario. Fig.~\ref{fig:convergence}(a) shows the population dynamics corresponding to Fig.~\ref{fig:path_dynamics} at $\varphi=1.17$, $T=4\,$K and increasing memory depths $n_c$ of the continuous reservoir. The left inset shows a zoom-in on the long term dynamics, while the right inset depicts the resulting dimension of the link index connecting the two tensor networks, demonstrating the here occurring exponential growth of inter-reservoir entanglement with increasing memory depths.
While the results are clearly not convergent at $n_c=2$ (dark blue line), the cases $n_c=3$ and $n_c=4$ (light blue and green lines) already are in very good agreement. Between the cases $n_c=4$ and $n_c=5$ (dashed orange line), no difference can be seen even at close range (see left inset), hence justifying the choice of $n_c=4$ employed in our calculations.
Fig.~\ref{fig:convergence}(b) shows the same dynamics calculated at increasing Schmidt value cutoff precisions $d_{cut}$ applied during the singular value decomposition~\cite{Schollwoeck2011}. Here, no differences between the cutoffs $d_{cut}=10^{-8}$ (dark blue line) and $d_{cut}=10^{-14}$ (dashed orange line) can be seen even at close range (see inset), underlining the convergence of our results with respect to the employed cutoff precision $d_{cut}=10^{-12}$ (green line). Hence, no crucial entanglement information has been truncated during the time evolution.

In addition, we investigate the numerical convergence of our results with respect to the time evolution step size $\Delta t$ to ensure that no errors occur during the Trotter decomposition. We first calculate the dynamics of both reservoirs independently for decreasing step sizes. Fig.~\ref{fig:stepsize}(a) shows the time evolution dynamics resulting from the time-discrete reservoir at $\varphi=1.17$ for decreasing step sizes $\Delta t$ (see Fig.~\ref{fig:lindblad}). Aside from minor differences during the initial time steps, the long term dynamics show very good agreement for all employed $\Delta t$, underlining the convergence of the feedback algorithm at the chosen step size $\Delta t=0.3\,$ps (light blue line).
Fig.~\ref{fig:stepsize}(b) shows the polarization dynamics imposed by the continuous reservoir [see Fig.~\ref{fig:mps_harmonic}(b)], calculated at $T=4\,$K and decreasing time evolution step sizes $\Delta t$. Again, the resulting long term dynamics at $\Delta t = \{0.3\,\mathrm{ps},0.15\,\mathrm{ps},0.075\,\mathrm{ps}\}$ are in good agreement.
Since both processes converge individually with respect to the time discretization, the combined setup can be expected to converge as well, since the involved time scales remain the same. Fig.~\ref{fig:stepsize}(c) shows the dynamics of the combined system coupled to both structured reservoirs for time discretizations $\Delta t = \{0.3\,\mathrm{ps},0.24\,\mathrm{ps}\}$, corresponding to $n_d=\{4,5\}$ feedback bins. Due to limited computational resources imposing restrictions on the memory depths of both reservoirs and thus the maximum achievable time discretization at a given feedback time $\tau$, the presented calculations are performed until $t=10\,$ps, where they exhibit good agreement with each other.
Lastly, in Fig.~\ref{fig:stepsize}(d) we verify the convergence of the numerical implementation with respect to the order of the series expansion performed for the Liouvillian $\mathcal{L}(t)$ [see Eq.~\eqref{eq:liou_evo}]. While the resulting dynamics show minor variations between the eighth and ninth order series expansions on a close scale (blue and green lines, see inset), no difference can be observed when comparing the dynamics resulting from ninth and tenth order expansions (dashed orange line). In conclusion, the employed tenth order series expansion of $\mathcal{L}(t)$ ensures convergent results as well.


\begin{acknowledgments}
The authors acknowledge support from the Deutsche Forschungsgemeinschaft (DFG) through SFB 910 project B1 (Project No. 163436311).
\end{acknowledgments}


\begin{thebibliography}{67}%
\makeatletter
\providecommand \@ifxundefined [1]{%
 \@ifx{#1\undefined}
}%
\providecommand \@ifnum [1]{%
 \ifnum #1\expandafter \@firstoftwo
 \else \expandafter \@secondoftwo
 \fi
}%
\providecommand \@ifx [1]{%
 \ifx #1\expandafter \@firstoftwo
 \else \expandafter \@secondoftwo
 \fi
}%
\providecommand \natexlab [1]{#1}%
\providecommand \enquote  [1]{``#1''}%
\providecommand \bibnamefont  [1]{#1}%
\providecommand \bibfnamefont [1]{#1}%
\providecommand \citenamefont [1]{#1}%
\providecommand \href@noop [0]{\@secondoftwo}%
\providecommand \href [0]{\begingroup \@sanitize@url \@href}%
\providecommand \@href[1]{\@@startlink{#1}\@@href}%
\providecommand \@@href[1]{\endgroup#1\@@endlink}%
\providecommand \@sanitize@url [0]{\catcode `\\12\catcode `\$12\catcode
  `\&12\catcode `\#12\catcode `\^12\catcode `\_12\catcode `\%12\relax}%
\providecommand \@@startlink[1]{}%
\providecommand \@@endlink[0]{}%
\providecommand \url  [0]{\begingroup\@sanitize@url \@url }%
\providecommand \@url [1]{\endgroup\@href {#1}{\urlprefix }}%
\providecommand \urlprefix  [0]{URL }%
\providecommand \Eprint [0]{\href }%
\providecommand \doibase [0]{http://dx.doi.org/}%
\providecommand \selectlanguage [0]{\@gobble}%
\providecommand \bibinfo  [0]{\@secondoftwo}%
\providecommand \bibfield  [0]{\@secondoftwo}%
\providecommand \translation [1]{[#1]}%
\providecommand \BibitemOpen [0]{}%
\providecommand \bibitemStop [0]{}%
\providecommand \bibitemNoStop [0]{.\EOS\space}%
\providecommand \EOS [0]{\spacefactor3000\relax}%
\providecommand \BibitemShut  [1]{\csname bibitem#1\endcsname}%
\let\auto@bib@innerbib\@empty
\bibitem [{\citenamefont {Schröder}\ \emph {et~al.}(2019)\citenamefont
  {Schröder}, \citenamefont {Turban}, \citenamefont {Musser}, \citenamefont
  {Hine},\ and\ \citenamefont {Chin}}]{Schroeder2019}%
  \BibitemOpen
  \bibfield  {author} {\bibinfo {author} {\bibfnamefont {F.~A. Y.~N.}\
  \bibnamefont {Schröder}}, \bibinfo {author} {\bibfnamefont {D.~H.~P.}\
  \bibnamefont {Turban}}, \bibinfo {author} {\bibfnamefont {A.~J.}\
  \bibnamefont {Musser}}, \bibinfo {author} {\bibfnamefont {N.~D.~M.}\
  \bibnamefont {Hine}}, \ and\ \bibinfo {author} {\bibfnamefont {A.~W.}\
  \bibnamefont {Chin}},\ }\href@noop {} {\bibfield  {journal} {\bibinfo
  {journal} {Nature Communications}\ }\textbf {\bibinfo {volume} {10}}
  (\bibinfo {year} {2019})}\BibitemShut {NoStop}%
\bibitem [{\citenamefont {Luchnikov}\ \emph {et~al.}(2019)\citenamefont
  {Luchnikov}, \citenamefont {Vintskevich}, \citenamefont {Ouerdane},\ and\
  \citenamefont {Filippov}}]{Luchnikov2019}%
  \BibitemOpen
  \bibfield  {author} {\bibinfo {author} {\bibfnamefont {I.~A.}\ \bibnamefont
  {Luchnikov}}, \bibinfo {author} {\bibfnamefont {S.~V.}\ \bibnamefont
  {Vintskevich}}, \bibinfo {author} {\bibfnamefont {H.}~\bibnamefont
  {Ouerdane}}, \ and\ \bibinfo {author} {\bibfnamefont {S.~N.}\ \bibnamefont
  {Filippov}},\ }\href {\doibase 10.1103/PhysRevLett.122.160401} {\bibfield
  {journal} {\bibinfo  {journal} {Phys. Rev. Lett.}\ }\textbf {\bibinfo
  {volume} {122}},\ \bibinfo {pages} {160401} (\bibinfo {year}
  {2019})}\BibitemShut {NoStop}%
\bibitem [{\citenamefont {Luchnikov}\ \emph {et~al.}(2020)\citenamefont
  {Luchnikov}, \citenamefont {Vintskevich}, \citenamefont {Grigoriev},\ and\
  \citenamefont {Filippov}}]{Luchnikov2020}%
  \BibitemOpen
  \bibfield  {author} {\bibinfo {author} {\bibfnamefont {I.~A.}\ \bibnamefont
  {Luchnikov}}, \bibinfo {author} {\bibfnamefont {S.~V.}\ \bibnamefont
  {Vintskevich}}, \bibinfo {author} {\bibfnamefont {D.~A.}\ \bibnamefont
  {Grigoriev}}, \ and\ \bibinfo {author} {\bibfnamefont {S.~N.}\ \bibnamefont
  {Filippov}},\ }\href {\doibase 10.1103/PhysRevLett.124.140502} {\bibfield
  {journal} {\bibinfo  {journal} {Phys. Rev. Lett.}\ }\textbf {\bibinfo
  {volume} {124}},\ \bibinfo {pages} {140502} (\bibinfo {year}
  {2020})}\BibitemShut {NoStop}%
\bibitem [{\citenamefont {Kuhn}\ and\ \citenamefont
  {Richter}(2019)}]{Kuhn2019}%
  \BibitemOpen
  \bibfield  {author} {\bibinfo {author} {\bibfnamefont {S.~C.}\ \bibnamefont
  {Kuhn}}\ and\ \bibinfo {author} {\bibfnamefont {M.}~\bibnamefont {Richter}},\
  }\href {\doibase 10.1103/PhysRevB.99.241301} {\bibfield  {journal} {\bibinfo
  {journal} {Phys. Rev. B}\ }\textbf {\bibinfo {volume} {99}},\ \bibinfo
  {pages} {241301} (\bibinfo {year} {2019})}\BibitemShut {NoStop}%
\bibitem [{\citenamefont {Breuer}\ \emph {et~al.}(2016)\citenamefont {Breuer},
  \citenamefont {Laine}, \citenamefont {Piilo},\ and\ \citenamefont
  {Vacchini}}]{Breuer2016}%
  \BibitemOpen
  \bibfield  {author} {\bibinfo {author} {\bibfnamefont {H.-P.}\ \bibnamefont
  {Breuer}}, \bibinfo {author} {\bibfnamefont {E.-M.}\ \bibnamefont {Laine}},
  \bibinfo {author} {\bibfnamefont {J.}~\bibnamefont {Piilo}}, \ and\ \bibinfo
  {author} {\bibfnamefont {B.}~\bibnamefont {Vacchini}},\ }\href {\doibase
  10.1103/RevModPhys.88.021002} {\bibfield  {journal} {\bibinfo  {journal}
  {Rev. Mod. Phys.}\ }\textbf {\bibinfo {volume} {88}},\ \bibinfo {pages}
  {021002} (\bibinfo {year} {2016})}\BibitemShut {NoStop}%
\bibitem [{\citenamefont {de~Vega}\ and\ \citenamefont
  {Alonso}(2017)}]{Vega2017}%
  \BibitemOpen
  \bibfield  {author} {\bibinfo {author} {\bibfnamefont {I.}~\bibnamefont
  {de~Vega}}\ and\ \bibinfo {author} {\bibfnamefont {D.}~\bibnamefont
  {Alonso}},\ }\href {\doibase 10.1103/RevModPhys.89.015001} {\bibfield
  {journal} {\bibinfo  {journal} {Rev. Mod. Phys.}\ }\textbf {\bibinfo {volume}
  {89}},\ \bibinfo {pages} {015001} (\bibinfo {year} {2017})}\BibitemShut
  {NoStop}%
\bibitem [{\citenamefont {Vagov}\ \emph
  {et~al.}(2011{\natexlab{a}})\citenamefont {Vagov}, \citenamefont {Croitoru},
  \citenamefont {Axt}, \citenamefont {Machnikowski},\ and\ \citenamefont
  {Kuhn}}]{Vagov2011pssb}%
  \BibitemOpen
  \bibfield  {author} {\bibinfo {author} {\bibfnamefont {A.}~\bibnamefont
  {Vagov}}, \bibinfo {author} {\bibfnamefont {M.~D.}\ \bibnamefont {Croitoru}},
  \bibinfo {author} {\bibfnamefont {V.~M.}\ \bibnamefont {Axt}}, \bibinfo
  {author} {\bibfnamefont {P.}~\bibnamefont {Machnikowski}}, \ and\ \bibinfo
  {author} {\bibfnamefont {T.}~\bibnamefont {Kuhn}},\ }\href {\doibase
  10.1002/pssb.201000842} {\bibfield  {journal} {\bibinfo  {journal} {physica
  status solidi (b)}\ }\textbf {\bibinfo {volume} {248}},\ \bibinfo {pages}
  {839} (\bibinfo {year} {2011}{\natexlab{a}})}\BibitemShut {NoStop}%
\bibitem [{\citenamefont {Gl{\"a}ssl}\ \emph {et~al.}(2011)\citenamefont
  {Gl{\"a}ssl}, \citenamefont {Vagov}, \citenamefont {L{\"u}ker}, \citenamefont
  {Reiter}, \citenamefont {Croitoru}, \citenamefont {Machnikowski},
  \citenamefont {Axt},\ and\ \citenamefont {Kuhn}}]{Glaessl2011}%
  \BibitemOpen
  \bibfield  {author} {\bibinfo {author} {\bibfnamefont {M.}~\bibnamefont
  {Gl{\"a}ssl}}, \bibinfo {author} {\bibfnamefont {A.}~\bibnamefont {Vagov}},
  \bibinfo {author} {\bibfnamefont {S.}~\bibnamefont {L{\"u}ker}}, \bibinfo
  {author} {\bibfnamefont {D.~E.}\ \bibnamefont {Reiter}}, \bibinfo {author}
  {\bibfnamefont {M.~D.}\ \bibnamefont {Croitoru}}, \bibinfo {author}
  {\bibfnamefont {P.}~\bibnamefont {Machnikowski}}, \bibinfo {author}
  {\bibfnamefont {V.~M.}\ \bibnamefont {Axt}}, \ and\ \bibinfo {author}
  {\bibfnamefont {T.}~\bibnamefont {Kuhn}},\ }\href {\doibase
  10.1103/PhysRevB.84.195311} {\bibfield  {journal} {\bibinfo  {journal} {Phys.
  Rev. B}\ }\textbf {\bibinfo {volume} {84}},\ \bibinfo {pages} {195311}
  (\bibinfo {year} {2011})}\BibitemShut {NoStop}%
\bibitem [{\citenamefont {Feynman}\ and\ \citenamefont
  {Hibbs}(1965)}]{Feynman1965}%
  \BibitemOpen
  \bibfield  {author} {\bibinfo {author} {\bibfnamefont {R.~P.}\ \bibnamefont
  {Feynman}}\ and\ \bibinfo {author} {\bibfnamefont {A.~R.}\ \bibnamefont
  {Hibbs}},\ }\href@noop {} {\emph {\bibinfo {title} {Quantum Mechanics and
  Path Integrals}}}\ (\bibinfo  {publisher} {McGraw-Hill College},\ \bibinfo
  {year} {1965})\BibitemShut {NoStop}%
\bibitem [{\citenamefont {Weiss}(2011)}]{Weiss2011}%
  \BibitemOpen
  \bibfield  {author} {\bibinfo {author} {\bibfnamefont {U.}~\bibnamefont
  {Weiss}},\ }\href {\doibase 10.1142/8334} {\emph {\bibinfo {title} {Quantum
  Dissipative Systems}}}\ (\bibinfo  {publisher} {World Scientific},\ \bibinfo
  {year} {2011})\BibitemShut {NoStop}%
\bibitem [{\citenamefont {Caldeira}\ and\ \citenamefont
  {Leggett}(1983)}]{Caldeira1983}%
  \BibitemOpen
  \bibfield  {author} {\bibinfo {author} {\bibfnamefont {A.}~\bibnamefont
  {Caldeira}}\ and\ \bibinfo {author} {\bibfnamefont {A.}~\bibnamefont
  {Leggett}},\ } {\bibfield
  {journal} {\bibinfo  {journal} {Physica A: Statistical Mechanics and its
  Applications}\ }\textbf {\bibinfo {volume} {121}},\ \bibinfo {pages} {587}
  (\bibinfo {year} {1983})}\BibitemShut {NoStop}%
\bibitem [{\citenamefont {Vagov}\ \emph
  {et~al.}(2011{\natexlab{b}})\citenamefont {Vagov}, \citenamefont {Croitoru},
  \citenamefont {Gl{\"a}ssl}, \citenamefont {Axt},\ and\ \citenamefont
  {Kuhn}}]{Vagov2011}%
  \BibitemOpen
  \bibfield  {author} {\bibinfo {author} {\bibfnamefont {A.}~\bibnamefont
  {Vagov}}, \bibinfo {author} {\bibfnamefont {M.~D.}\ \bibnamefont {Croitoru}},
  \bibinfo {author} {\bibfnamefont {M.}~\bibnamefont {Gl{\"a}ssl}}, \bibinfo
  {author} {\bibfnamefont {V.~M.}\ \bibnamefont {Axt}}, \ and\ \bibinfo
  {author} {\bibfnamefont {T.}~\bibnamefont {Kuhn}},\ }\href {\doibase
  10.1103/PhysRevB.83.094303} {\bibfield  {journal} {\bibinfo  {journal} {Phys.
  Rev. B}\ }\textbf {\bibinfo {volume} {83}},\ \bibinfo {pages} {094303}
  (\bibinfo {year} {2011}{\natexlab{b}})}\BibitemShut {NoStop}%
\bibitem [{\citenamefont {Gl\"assl}\ \emph {et~al.}(2013)\citenamefont
  {Gl\"assl}, \citenamefont {Barth},\ and\ \citenamefont {Axt}}]{Glaessl2013}%
  \BibitemOpen
  \bibfield  {author} {\bibinfo {author} {\bibfnamefont {M.}~\bibnamefont
  {Gl\"assl}}, \bibinfo {author} {\bibfnamefont {A.~M.}\ \bibnamefont {Barth}},
  \ and\ \bibinfo {author} {\bibfnamefont {V.~M.}\ \bibnamefont {Axt}},\ }\href
  {\doibase 10.1103/PhysRevLett.110.147401} {\bibfield  {journal} {\bibinfo
  {journal} {Phys. Rev. Lett.}\ }\textbf {\bibinfo {volume} {110}},\ \bibinfo
  {pages} {147401} (\bibinfo {year} {2013})}\BibitemShut {NoStop}%
\bibitem [{\citenamefont {Barth}\ \emph {et~al.}(2016)\citenamefont {Barth},
  \citenamefont {Vagov},\ and\ \citenamefont {Axt}}]{Barth2016}%
  \BibitemOpen
  \bibfield  {author} {\bibinfo {author} {\bibfnamefont {A.~M.}\ \bibnamefont
  {Barth}}, \bibinfo {author} {\bibfnamefont {A.}~\bibnamefont {Vagov}}, \ and\
  \bibinfo {author} {\bibfnamefont {V.~M.}\ \bibnamefont {Axt}},\ }\href
  {\doibase 10.1103/PhysRevB.94.125439} {\bibfield  {journal} {\bibinfo
  {journal} {Phys. Rev. B}\ }\textbf {\bibinfo {volume} {94}},\ \bibinfo
  {pages} {125439} (\bibinfo {year} {2016})}\BibitemShut {NoStop}%
\bibitem [{\citenamefont {Cosacchi}\ \emph {et~al.}(2018)\citenamefont
  {Cosacchi}, \citenamefont {Cygorek}, \citenamefont {Ungar}, \citenamefont
  {Barth}, \citenamefont {Vagov},\ and\ \citenamefont {Axt}}]{Cosacchi2018}%
  \BibitemOpen
  \bibfield  {author} {\bibinfo {author} {\bibfnamefont {M.}~\bibnamefont
  {Cosacchi}}, \bibinfo {author} {\bibfnamefont {M.}~\bibnamefont {Cygorek}},
  \bibinfo {author} {\bibfnamefont {F.}~\bibnamefont {Ungar}}, \bibinfo
  {author} {\bibfnamefont {A.~M.}\ \bibnamefont {Barth}}, \bibinfo {author}
  {\bibfnamefont {A.}~\bibnamefont {Vagov}}, \ and\ \bibinfo {author}
  {\bibfnamefont {V.~M.}\ \bibnamefont {Axt}},\ }\href {\doibase
  10.1103/PhysRevB.98.125302} {\bibfield  {journal} {\bibinfo  {journal} {Phys.
  Rev. B}\ }\textbf {\bibinfo {volume} {98}},\ \bibinfo {pages} {125302}
  (\bibinfo {year} {2018})}\BibitemShut {NoStop}%
\bibitem [{\citenamefont {Makri}\ and\ \citenamefont
  {Makarov}(1995{\natexlab{a}})}]{Makri1995}%
  \BibitemOpen
  \bibfield  {author} {\bibinfo {author} {\bibfnamefont {N.}~\bibnamefont
  {Makri}}\ and\ \bibinfo {author} {\bibfnamefont {D.~E.}\ \bibnamefont
  {Makarov}},\ }\href {\doibase 10.1063/1.469508} {\bibfield  {journal}
  {\bibinfo  {journal} {The Journal of Chemical Physics}\ }\textbf {\bibinfo
  {volume} {102}},\ \bibinfo {pages} {4600} (\bibinfo {year}
  {1995}{\natexlab{a}})}\BibitemShut {NoStop}%
\bibitem [{\citenamefont {Makri}\ and\ \citenamefont
  {Makarov}(1995{\natexlab{b}})}]{Makri1995a}%
  \BibitemOpen
  \bibfield  {author} {\bibinfo {author} {\bibfnamefont {N.}~\bibnamefont
  {Makri}}\ and\ \bibinfo {author} {\bibfnamefont {D.~E.}\ \bibnamefont
  {Makarov}},\ }\href {\doibase 10.1063/1.469509} {\bibfield  {journal}
  {\bibinfo  {journal} {The Journal of Chemical Physics}\ }\textbf {\bibinfo
  {volume} {102}},\ \bibinfo {pages} {4611} (\bibinfo {year}
  {1995}{\natexlab{b}})}\BibitemShut {NoStop}%
\bibitem [{\citenamefont {Budini}(2018)}]{Budini2018}%
  \BibitemOpen
  \bibfield  {author} {\bibinfo {author} {\bibfnamefont {A.~A.}\ \bibnamefont
  {Budini}},\ }\href {\doibase 10.1103/PhysRevLett.121.240401} {\bibfield
  {journal} {\bibinfo  {journal} {Phys. Rev. Lett.}\ }\textbf {\bibinfo
  {volume} {121}},\ \bibinfo {pages} {240401} (\bibinfo {year}
  {2018})}\BibitemShut {NoStop}%
\bibitem [{\citenamefont {Taranto}\ \emph {et~al.}(2019)\citenamefont
  {Taranto}, \citenamefont {Pollock}, \citenamefont {Milz}, \citenamefont
  {Tomamichel},\ and\ \citenamefont {Modi}}]{Taranto2019}%
  \BibitemOpen
  \bibfield  {author} {\bibinfo {author} {\bibfnamefont {P.}~\bibnamefont
  {Taranto}}, \bibinfo {author} {\bibfnamefont {F.~A.}\ \bibnamefont
  {Pollock}}, \bibinfo {author} {\bibfnamefont {S.}~\bibnamefont {Milz}},
  \bibinfo {author} {\bibfnamefont {M.}~\bibnamefont {Tomamichel}}, \ and\
  \bibinfo {author} {\bibfnamefont {K.}~\bibnamefont {Modi}},\ }\href {\doibase
  10.1103/PhysRevLett.122.140401} {\bibfield  {journal} {\bibinfo  {journal}
  {Phys. Rev. Lett.}\ }\textbf {\bibinfo {volume} {122}},\ \bibinfo {pages}
  {140401} (\bibinfo {year} {2019})}\BibitemShut {NoStop}%
\bibitem [{\citenamefont {Pollock}\ \emph
  {et~al.}(2018{\natexlab{a}})\citenamefont {Pollock}, \citenamefont
  {Rodr\'{\i}guez-Rosario}, \citenamefont {Frauenheim}, \citenamefont
  {Paternostro},\ and\ \citenamefont {Modi}}]{Pollock2018}%
  \BibitemOpen
  \bibfield  {author} {\bibinfo {author} {\bibfnamefont {F.~A.}\ \bibnamefont
  {Pollock}}, \bibinfo {author} {\bibfnamefont {C.}~\bibnamefont
  {Rodr\'{\i}guez-Rosario}}, \bibinfo {author} {\bibfnamefont {T.}~\bibnamefont
  {Frauenheim}}, \bibinfo {author} {\bibfnamefont {M.}~\bibnamefont
  {Paternostro}}, \ and\ \bibinfo {author} {\bibfnamefont {K.}~\bibnamefont
  {Modi}},\ }\href {\doibase 10.1103/PhysRevLett.120.040405} {\bibfield
  {journal} {\bibinfo  {journal} {Phys. Rev. Lett.}\ }\textbf {\bibinfo
  {volume} {120}},\ \bibinfo {pages} {040405} (\bibinfo {year}
  {2018}{\natexlab{a}})}\BibitemShut {NoStop}%
\bibitem [{\citenamefont {Pollock}\ \emph
  {et~al.}(2018{\natexlab{b}})\citenamefont {Pollock}, \citenamefont
  {Rodr\'{\i}guez-Rosario}, \citenamefont {Frauenheim}, \citenamefont
  {Paternostro},\ and\ \citenamefont {Modi}}]{Pollock2018a}%
  \BibitemOpen
  \bibfield  {author} {\bibinfo {author} {\bibfnamefont {F.~A.}\ \bibnamefont
  {Pollock}}, \bibinfo {author} {\bibfnamefont {C.}~\bibnamefont
  {Rodr\'{\i}guez-Rosario}}, \bibinfo {author} {\bibfnamefont {T.}~\bibnamefont
  {Frauenheim}}, \bibinfo {author} {\bibfnamefont {M.}~\bibnamefont
  {Paternostro}}, \ and\ \bibinfo {author} {\bibfnamefont {K.}~\bibnamefont
  {Modi}},\ }\href {\doibase 10.1103/PhysRevA.97.012127} {\bibfield  {journal}
  {\bibinfo  {journal} {Phys. Rev. A}\ }\textbf {\bibinfo {volume} {97}},\
  \bibinfo {pages} {012127} (\bibinfo {year} {2018}{\natexlab{b}})}\BibitemShut
  {NoStop}%
\bibitem [{\citenamefont {Li}\ \emph {et~al.}(2018)\citenamefont {Li},
  \citenamefont {Hall},\ and\ \citenamefont {Wiseman}}]{Li2018}%
  \BibitemOpen
  \bibfield  {author} {\bibinfo {author} {\bibfnamefont {L.}~\bibnamefont
  {Li}}, \bibinfo {author} {\bibfnamefont {M.~J.}\ \bibnamefont {Hall}}, \ and\
  \bibinfo {author} {\bibfnamefont {H.~M.}\ \bibnamefont {Wiseman}},\ }\href
  {\doibase 10.1016/j.physrep.2018.07.001} {\bibfield  {journal} {\bibinfo
  {journal} {Physics Reports}\ }\textbf {\bibinfo {volume} {759}},\ \bibinfo
  {pages} {1} (\bibinfo {year} {2018})}\BibitemShut {NoStop}%
\bibitem [{\citenamefont {Li}\ \emph {et~al.}(2019)\citenamefont {Li},
  \citenamefont {Guo},\ and\ \citenamefont {Piilo}}]{Li2019}%
  \BibitemOpen
  \bibfield  {author} {\bibinfo {author} {\bibfnamefont {C.-F.}\ \bibnamefont
  {Li}}, \bibinfo {author} {\bibfnamefont {G.-C.}\ \bibnamefont {Guo}}, \ and\
  \bibinfo {author} {\bibfnamefont {J.}~\bibnamefont {Piilo}},\ }\href
  {\doibase 10.1209/0295-5075/127/50001} {\bibfield  {journal} {\bibinfo
  {journal} {{EPL} (Europhysics Letters)}\ }\textbf {\bibinfo {volume} {127}},\
  \bibinfo {pages} {50001} (\bibinfo {year} {2019})}\BibitemShut {NoStop}%
\bibitem [{\citenamefont {del Pino}\ \emph {et~al.}(2018)\citenamefont {del
  Pino}, \citenamefont {Schr\"oder}, \citenamefont {Chin}, \citenamefont
  {Feist},\ and\ \citenamefont {Garcia-Vidal}}]{delPino2018}%
  \BibitemOpen
  \bibfield  {author} {\bibinfo {author} {\bibfnamefont {J.}~\bibnamefont {del
  Pino}}, \bibinfo {author} {\bibfnamefont {F.~A. Y.~N.}\ \bibnamefont
  {Schr\"oder}}, \bibinfo {author} {\bibfnamefont {A.~W.}\ \bibnamefont
  {Chin}}, \bibinfo {author} {\bibfnamefont {J.}~\bibnamefont {Feist}}, \ and\
  \bibinfo {author} {\bibfnamefont {F.~J.}\ \bibnamefont {Garcia-Vidal}},\
  }\href {\doibase 10.1103/PhysRevLett.121.227401} {\bibfield  {journal}
  {\bibinfo  {journal} {Phys. Rev. Lett.}\ }\textbf {\bibinfo {volume} {121}},\
  \bibinfo {pages} {227401} (\bibinfo {year} {2018})}\BibitemShut {NoStop}%
\bibitem [{\citenamefont {Regidor}\ \emph {et~al.}(2020)\citenamefont
  {Regidor}, \citenamefont {Crowder}, \citenamefont {Carmichael},\ and\
  \citenamefont {Hughes}}]{Regidor2020}%
  \BibitemOpen
  \bibfield  {author} {\bibinfo {author} {\bibfnamefont {S.~A.}\ \bibnamefont
  {Regidor}}, \bibinfo {author} {\bibfnamefont {G.}~\bibnamefont {Crowder}},
  \bibinfo {author} {\bibfnamefont {H.}~\bibnamefont {Carmichael}}, \ and\
  \bibinfo {author} {\bibfnamefont {S.}~\bibnamefont {Hughes}},\ }\href@noop {}
  {\enquote {\bibinfo {title} {Modelling quantum light-matter interactions in
  waveguide-qed with retardation and a time-delayed feedback: matrix product
  states versus a space-discretized waveguide model},}\ } (\bibinfo {year}
  {2020}),\ \Eprint {http://arxiv.org/abs/2011.12205} {arXiv:2011.12205}
  \BibitemShut {NoStop}%
\bibitem [{\citenamefont {Denning}\ \emph {et~al.}(2020)\citenamefont
  {Denning}, \citenamefont {Bundgaard-Nielsen},\ and\ \citenamefont
  {M\o{}rk}}]{DenningBundgaard2020}%
  \BibitemOpen
  \bibfield  {author} {\bibinfo {author} {\bibfnamefont {E.~V.}\ \bibnamefont
  {Denning}}, \bibinfo {author} {\bibfnamefont {M.}~\bibnamefont
  {Bundgaard-Nielsen}}, \ and\ \bibinfo {author} {\bibfnamefont
  {J.}~\bibnamefont {M\o{}rk}},\ }\href {\doibase 10.1103/PhysRevB.102.235303}
  {\bibfield  {journal} {\bibinfo  {journal} {Phys. Rev. B}\ }\textbf {\bibinfo
  {volume} {102}},\ \bibinfo {pages} {235303} (\bibinfo {year}
  {2020})}\BibitemShut {NoStop}%
\bibitem [{\citenamefont {Pichler}\ and\ \citenamefont
  {Zoller}(2016)}]{Pichler2016}%
  \BibitemOpen
  \bibfield  {author} {\bibinfo {author} {\bibfnamefont {H.}~\bibnamefont
  {Pichler}}\ and\ \bibinfo {author} {\bibfnamefont {P.}~\bibnamefont
  {Zoller}},\ }\href {\doibase 10.1103/PhysRevLett.116.093601} {\bibfield
  {journal} {\bibinfo  {journal} {Phys. Rev. Lett.}\ }\textbf {\bibinfo
  {volume} {116}},\ \bibinfo {pages} {093601} (\bibinfo {year}
  {2016})}\BibitemShut {NoStop}%
\bibitem [{\citenamefont {Guimond}\ \emph {et~al.}(2016)\citenamefont
  {Guimond}, \citenamefont {Pichler}, \citenamefont {Rauschenbeutel},\ and\
  \citenamefont {Zoller}}]{Guimond2016}%
  \BibitemOpen
  \bibfield  {author} {\bibinfo {author} {\bibfnamefont {P.-O.}\ \bibnamefont
  {Guimond}}, \bibinfo {author} {\bibfnamefont {H.}~\bibnamefont {Pichler}},
  \bibinfo {author} {\bibfnamefont {A.}~\bibnamefont {Rauschenbeutel}}, \ and\
  \bibinfo {author} {\bibfnamefont {P.}~\bibnamefont {Zoller}},\ }\href
  {\doibase 10.1103/PhysRevA.94.033829} {\bibfield  {journal} {\bibinfo
  {journal} {Phys. Rev. A}\ }\textbf {\bibinfo {volume} {94}},\ \bibinfo
  {pages} {033829} (\bibinfo {year} {2016})}\BibitemShut {NoStop}%
\bibitem [{\citenamefont {Guimond}\ \emph {et~al.}(2017)\citenamefont
  {Guimond}, \citenamefont {Pletyukhov}, \citenamefont {Pichler},\ and\
  \citenamefont {Zoller}}]{Guimond2017}%
  \BibitemOpen
  \bibfield  {author} {\bibinfo {author} {\bibfnamefont {P.-O.}\ \bibnamefont
  {Guimond}}, \bibinfo {author} {\bibfnamefont {M.}~\bibnamefont {Pletyukhov}},
  \bibinfo {author} {\bibfnamefont {H.}~\bibnamefont {Pichler}}, \ and\
  \bibinfo {author} {\bibfnamefont {P.}~\bibnamefont {Zoller}},\ }\href
  {\doibase 10.1088/2058-9565/aa7f03} {\bibfield  {journal} {\bibinfo
  {journal} {Quantum Science and Technology}\ }\textbf {\bibinfo {volume}
  {2}},\ \bibinfo {pages} {044012} (\bibinfo {year} {2017})}\BibitemShut
  {NoStop}%
\bibitem [{\citenamefont {Strathearn}\ \emph {et~al.}(2017)\citenamefont
  {Strathearn}, \citenamefont {Lovett},\ and\ \citenamefont
  {Kirton}}]{Strathearn2017}%
  \BibitemOpen
  \bibfield  {author} {\bibinfo {author} {\bibfnamefont {A.}~\bibnamefont
  {Strathearn}}, \bibinfo {author} {\bibfnamefont {B.~W.}\ \bibnamefont
  {Lovett}}, \ and\ \bibinfo {author} {\bibfnamefont {P.}~\bibnamefont
  {Kirton}},\ }\href@noop {} {\bibfield  {journal} {\bibinfo  {journal} {New
  Journal of Physics}\ }\textbf {\bibinfo {volume} {19}},\ \bibinfo {pages}
  {093009} (\bibinfo {year} {2017})}\BibitemShut {NoStop}%
\bibitem [{\citenamefont {Strathearn}\ \emph {et~al.}(2018)\citenamefont
  {Strathearn}, \citenamefont {Kirton}, \citenamefont {Kilda}, \citenamefont
  {Keeling},\ and\ \citenamefont {Lovett}}]{Strathearn2018}%
  \BibitemOpen
  \bibfield  {author} {\bibinfo {author} {\bibfnamefont {A.}~\bibnamefont
  {Strathearn}}, \bibinfo {author} {\bibfnamefont {P.}~\bibnamefont {Kirton}},
  \bibinfo {author} {\bibfnamefont {D.}~\bibnamefont {Kilda}}, \bibinfo
  {author} {\bibfnamefont {J.}~\bibnamefont {Keeling}}, \ and\ \bibinfo
  {author} {\bibfnamefont {B.~W.}\ \bibnamefont {Lovett}},\ }\href@noop {}
  {\bibfield  {journal} {\bibinfo  {journal} {Nature Communications}\ }\textbf
  {\bibinfo {volume} {9}},\ \bibinfo {pages} {3322} (\bibinfo {year}
  {2018})}\BibitemShut {NoStop}%
\bibitem [{\citenamefont {J\o{}rgensen}\ and\ \citenamefont
  {Pollock}(2019)}]{Jorgensen2019}%
  \BibitemOpen
  \bibfield  {author} {\bibinfo {author} {\bibfnamefont {M.~R.}\ \bibnamefont
  {J\o{}rgensen}}\ and\ \bibinfo {author} {\bibfnamefont {F.~A.}\ \bibnamefont
  {Pollock}},\ }\href {\doibase 10.1103/PhysRevLett.123.240602} {\bibfield
  {journal} {\bibinfo  {journal} {Phys. Rev. Lett.}\ }\textbf {\bibinfo
  {volume} {123}},\ \bibinfo {pages} {240602} (\bibinfo {year}
  {2019})}\BibitemShut {NoStop}%
\bibitem [{\citenamefont {Gribben}\ \emph {et~al.}(2020)\citenamefont
  {Gribben}, \citenamefont {Strathearn}, \citenamefont {Iles-Smith},
  \citenamefont {Kilda}, \citenamefont {Nazir}, \citenamefont {Lovett},\ and\
  \citenamefont {Kirton}}]{Gribben2020}%
  \BibitemOpen
  \bibfield  {author} {\bibinfo {author} {\bibfnamefont {D.}~\bibnamefont
  {Gribben}}, \bibinfo {author} {\bibfnamefont {A.}~\bibnamefont {Strathearn}},
  \bibinfo {author} {\bibfnamefont {J.}~\bibnamefont {Iles-Smith}}, \bibinfo
  {author} {\bibfnamefont {D.}~\bibnamefont {Kilda}}, \bibinfo {author}
  {\bibfnamefont {A.}~\bibnamefont {Nazir}}, \bibinfo {author} {\bibfnamefont
  {B.~W.}\ \bibnamefont {Lovett}}, \ and\ \bibinfo {author} {\bibfnamefont
  {P.}~\bibnamefont {Kirton}},\ }\href {\doibase
  10.1103/PhysRevResearch.2.013265} {\bibfield  {journal} {\bibinfo  {journal}
  {Phys. Rev. Research}\ }\textbf {\bibinfo {volume} {2}},\ \bibinfo {pages}
  {013265} (\bibinfo {year} {2020})}\BibitemShut {NoStop}%
\bibitem [{\citenamefont {Strathearn}(2020)}]{Strathearn2020thesis}%
  \BibitemOpen
  \bibfield  {author} {\bibinfo {author} {\bibfnamefont {A.}~\bibnamefont
  {Strathearn}},\ } {\emph {\bibinfo
  {title} {Modelling Non-Markovian Quantum Systems Using Tensor Networks}}}\
  (\bibinfo  {publisher} {Springer International Publishing},\ \bibinfo {year}
  {2020})\BibitemShut {NoStop}%
\bibitem [{\citenamefont {Kshetrimayum}\ \emph {et~al.}(2019)\citenamefont
  {Kshetrimayum}, \citenamefont {Rizzi}, \citenamefont {Eisert},\ and\
  \citenamefont {Or\'us}}]{Eisert2019}%
  \BibitemOpen
  \bibfield  {author} {\bibinfo {author} {\bibfnamefont {A.}~\bibnamefont
  {Kshetrimayum}}, \bibinfo {author} {\bibfnamefont {M.}~\bibnamefont {Rizzi}},
  \bibinfo {author} {\bibfnamefont {J.}~\bibnamefont {Eisert}}, \ and\ \bibinfo
  {author} {\bibfnamefont {R.}~\bibnamefont {Or\'us}},\ }\href {\doibase
  10.1103/PhysRevLett.122.070502} {\bibfield  {journal} {\bibinfo  {journal}
  {Phys. Rev. Lett.}\ }\textbf {\bibinfo {volume} {122}},\ \bibinfo {pages}
  {070502} (\bibinfo {year} {2019})}\BibitemShut {NoStop}%
\bibitem [{\citenamefont {Tamascelli}\ \emph {et~al.}(2019)\citenamefont
  {Tamascelli}, \citenamefont {Smirne}, \citenamefont {Lim}, \citenamefont
  {Huelga},\ and\ \citenamefont {Plenio}}]{Tamascelli2019}%
  \BibitemOpen
  \bibfield  {author} {\bibinfo {author} {\bibfnamefont {D.}~\bibnamefont
  {Tamascelli}}, \bibinfo {author} {\bibfnamefont {A.}~\bibnamefont {Smirne}},
  \bibinfo {author} {\bibfnamefont {J.}~\bibnamefont {Lim}}, \bibinfo {author}
  {\bibfnamefont {S.~F.}\ \bibnamefont {Huelga}}, \ and\ \bibinfo {author}
  {\bibfnamefont {M.~B.}\ \bibnamefont {Plenio}},\ }\href {\doibase
  10.1103/PhysRevLett.123.090402} {\bibfield  {journal} {\bibinfo  {journal}
  {Phys. Rev. Lett.}\ }\textbf {\bibinfo {volume} {123}},\ \bibinfo {pages}
  {090402} (\bibinfo {year} {2019})}\BibitemShut {NoStop}%
\bibitem [{\citenamefont {Finsterhölzl}\ \emph {et~al.}(2020)\citenamefont
  {Finsterhölzl}, \citenamefont {Katzer}, \citenamefont {Knorr},\ and\
  \citenamefont {Carmele}}]{Finsterhoelzl2020}%
  \BibitemOpen
  \bibfield  {author} {\bibinfo {author} {\bibfnamefont {R.}~\bibnamefont
  {Finsterhölzl}}, \bibinfo {author} {\bibfnamefont {M.}~\bibnamefont
  {Katzer}}, \bibinfo {author} {\bibfnamefont {A.}~\bibnamefont {Knorr}}, \
  and\ \bibinfo {author} {\bibfnamefont {A.}~\bibnamefont {Carmele}},\ }\href
  {\doibase 10.3390/e22090984} {\bibfield  {journal} {\bibinfo  {journal}
  {Entropy}\ }\textbf {\bibinfo {volume} {22}},\ \bibinfo {pages} {984}
  (\bibinfo {year} {2020})}\BibitemShut {NoStop}%
\bibitem [{\citenamefont {Finsterh\"olzl}\ \emph {et~al.}(2020)\citenamefont
  {Finsterh\"olzl}, \citenamefont {Katzer},\ and\ \citenamefont
  {Carmele}}]{FinsterhoelzlCarmele2020}%
  \BibitemOpen
  \bibfield  {author} {\bibinfo {author} {\bibfnamefont {R.}~\bibnamefont
  {Finsterh\"olzl}}, \bibinfo {author} {\bibfnamefont {M.}~\bibnamefont
  {Katzer}}, \ and\ \bibinfo {author} {\bibfnamefont {A.}~\bibnamefont
  {Carmele}},\ }\href {\doibase 10.1103/PhysRevB.102.174309} {\bibfield
  {journal} {\bibinfo  {journal} {Phys. Rev. B}\ }\textbf {\bibinfo {volume}
  {102}},\ \bibinfo {pages} {174309} (\bibinfo {year} {2020})}\BibitemShut
  {NoStop}%
\bibitem [{\citenamefont {Calaj\'o}\ \emph
  {et~al.}(2019{\natexlab{a}})\citenamefont {Calaj\'o}, \citenamefont {Fang},
  \citenamefont {Baranger},\ and\ \citenamefont {Ciccarello}}]{Calajo2019}%
  \BibitemOpen
  \bibfield  {author} {\bibinfo {author} {\bibfnamefont {G.}~\bibnamefont
  {Calaj\'o}}, \bibinfo {author} {\bibfnamefont {Y.-L.~L.}\ \bibnamefont
  {Fang}}, \bibinfo {author} {\bibfnamefont {H.~U.}\ \bibnamefont {Baranger}},
  \ and\ \bibinfo {author} {\bibfnamefont {F.}~\bibnamefont {Ciccarello}},\
  }\href {\doibase 10.1103/PhysRevLett.122.073601} {\bibfield  {journal}
  {\bibinfo  {journal} {Phys. Rev. Lett.}\ }\textbf {\bibinfo {volume} {122}},\
  \bibinfo {pages} {073601} (\bibinfo {year} {2019}{\natexlab{a}})}\BibitemShut
  {NoStop}%
\bibitem [{\citenamefont {Crowder}\ \emph {et~al.}(2020)\citenamefont
  {Crowder}, \citenamefont {Carmichael},\ and\ \citenamefont
  {Hughes}}]{Crowder2020}%
  \BibitemOpen
  \bibfield  {author} {\bibinfo {author} {\bibfnamefont {G.}~\bibnamefont
  {Crowder}}, \bibinfo {author} {\bibfnamefont {H.}~\bibnamefont {Carmichael}},
  \ and\ \bibinfo {author} {\bibfnamefont {S.}~\bibnamefont {Hughes}},\ }\href
  {\doibase 10.1103/PhysRevA.101.023807} {\bibfield  {journal} {\bibinfo
  {journal} {Phys. Rev. A}\ }\textbf {\bibinfo {volume} {101}},\ \bibinfo
  {pages} {023807} (\bibinfo {year} {2020})}\BibitemShut {NoStop}%
\bibitem [{\citenamefont {Wang}\ \emph {et~al.}(2020)\citenamefont {Wang},
  \citenamefont {Jaako}, \citenamefont {Kirton},\ and\ \citenamefont
  {Rabl}}]{WangRabl2020}%
  \BibitemOpen
  \bibfield  {author} {\bibinfo {author} {\bibfnamefont {Z.}~\bibnamefont
  {Wang}}, \bibinfo {author} {\bibfnamefont {T.}~\bibnamefont {Jaako}},
  \bibinfo {author} {\bibfnamefont {P.}~\bibnamefont {Kirton}}, \ and\ \bibinfo
  {author} {\bibfnamefont {P.}~\bibnamefont {Rabl}},\ }\href {\doibase
  10.1103/PhysRevLett.124.213601} {\bibfield  {journal} {\bibinfo  {journal}
  {Phys. Rev. Lett.}\ }\textbf {\bibinfo {volume} {124}},\ \bibinfo {pages}
  {213601} (\bibinfo {year} {2020})}\BibitemShut {NoStop}%
\bibitem [{\citenamefont {Carmele}\ \emph
  {et~al.}(2020{\natexlab{a}})\citenamefont {Carmele}, \citenamefont {Nemet},
  \citenamefont {Canela},\ and\ \citenamefont {Parkins}}]{CarmeleNemet2020}%
  \BibitemOpen
  \bibfield  {author} {\bibinfo {author} {\bibfnamefont {A.}~\bibnamefont
  {Carmele}}, \bibinfo {author} {\bibfnamefont {N.}~\bibnamefont {Nemet}},
  \bibinfo {author} {\bibfnamefont {V.}~\bibnamefont {Canela}}, \ and\ \bibinfo
  {author} {\bibfnamefont {S.}~\bibnamefont {Parkins}},\ }\href {\doibase
  10.1103/PhysRevResearch.2.013238} {\bibfield  {journal} {\bibinfo  {journal}
  {Phys. Rev. Research}\ }\textbf {\bibinfo {volume} {2}},\ \bibinfo {pages}
  {013238} (\bibinfo {year} {2020}{\natexlab{a}})}\BibitemShut {NoStop}%
\bibitem [{\citenamefont {Leggett}\ \emph {et~al.}(1987)\citenamefont
  {Leggett}, \citenamefont {Chakravarty}, \citenamefont {Dorsey}, \citenamefont
  {Fisher}, \citenamefont {Garg},\ and\ \citenamefont {Zwerger}}]{Leggett1987}%
  \BibitemOpen
  \bibfield  {author} {\bibinfo {author} {\bibfnamefont {A.~J.}\ \bibnamefont
  {Leggett}}, \bibinfo {author} {\bibfnamefont {S.}~\bibnamefont
  {Chakravarty}}, \bibinfo {author} {\bibfnamefont {A.~T.}\ \bibnamefont
  {Dorsey}}, \bibinfo {author} {\bibfnamefont {M.~P.~A.}\ \bibnamefont
  {Fisher}}, \bibinfo {author} {\bibfnamefont {A.}~\bibnamefont {Garg}}, \ and\
  \bibinfo {author} {\bibfnamefont {W.}~\bibnamefont {Zwerger}},\ }\href
  {\doibase 10.1103/RevModPhys.59.1} {\bibfield  {journal} {\bibinfo  {journal}
  {Rev. Mod. Phys.}\ }\textbf {\bibinfo {volume} {59}},\ \bibinfo {pages} {1}
  (\bibinfo {year} {1987})}\BibitemShut {NoStop}%
\bibitem [{\citenamefont {Breuer}\ and\ \citenamefont
  {Petruccione}(2002)}]{Breuer2002}%
  \BibitemOpen
  \bibfield  {author} {\bibinfo {author} {\bibfnamefont {H.~P.}\ \bibnamefont
  {Breuer}}\ and\ \bibinfo {author} {\bibfnamefont {F.}~\bibnamefont
  {Petruccione}},\ }\href@noop {} {\emph {\bibinfo {title} {{The theory of open
  quantum systems}}}}\ (\bibinfo  {publisher} {Oxford University Press},\
  \bibinfo {year} {2002})\BibitemShut {NoStop}%
\bibitem [{\citenamefont {Mukamel}(1999)}]{Mukamel1999}%
  \BibitemOpen
  \bibfield  {author} {\bibinfo {author} {\bibfnamefont {S.}~\bibnamefont
  {Mukamel}},\ }\href@noop {} {\emph {\bibinfo {title} {Principles of nonlinear
  optical spectroscopy}}}\ (\bibinfo  {publisher} {Oxford University Press},\
  \bibinfo {year} {1999})\BibitemShut {NoStop}%
\bibitem [{\citenamefont {Schollw{\"o}ck}(2011)}]{Schollwoeck2011}%
  \BibitemOpen
  \bibfield  {author} {\bibinfo {author} {\bibfnamefont {U.}~\bibnamefont
  {Schollw{\"o}ck}},\ }\href {\doibase 10.1016/j.aop.2010.09.012} {\bibfield
  {journal} {\bibinfo  {journal} {Annals of Physics}\ }\textbf {\bibinfo
  {volume} {326}},\ \bibinfo {pages} {96} (\bibinfo {year} {2011})}\BibitemShut
  {NoStop}%
\bibitem [{\citenamefont {Mahan}(2000)}]{Mahan2000}%
  \BibitemOpen
  \bibfield  {author} {\bibinfo {author} {\bibfnamefont {G.~D.}\ \bibnamefont
  {Mahan}},\ }\href@noop {} {\emph {\bibinfo {title} {{Many-Particle
  Physics}}}}\ (\bibinfo  {publisher} {Kluwer Academic/Plenum},\ \bibinfo
  {year} {2000})\BibitemShut {NoStop}%
\bibitem [{Note1()}]{Note1}%
  \BibitemOpen
  \bibinfo {note} {As generic coupling, we choose the acoustic bulk phonon
  coupling element of GaAs, given by $g_{\protect \bm {q}}^{ii} = \protect
  \sqrt { \hbar q/(2 \rho c_s )} D_i \protect \qopname \relax o{exp}[ - \hbar
  q^2/(4 m_i \omega _i) ]$, resulting in a transition coupling element
  $g_{\protect \bm {q}} := g_{\protect \bm {q}}^{22} - g_{\protect \bm
  {q}}^{11}$~\cite {Carmele2019, Foerstner2003, *Foerstner2003pssb,
  CarmeleMilde2013}. Here, $D_i$ are deformation potentials, $m_i$ denote
  effective masses, $\hbar \omega _i$ refer to the confinement energies and
  $\rho $ is the mass density of GaAs, respectively.}\BibitemShut {Stop}%
\bibitem [{\citenamefont {Carmele}\ and\ \citenamefont
  {Reitzenstein}(2019)}]{Carmele2019}%
  \BibitemOpen
  \bibfield  {author} {\bibinfo {author} {\bibfnamefont {A.}~\bibnamefont
  {Carmele}}\ and\ \bibinfo {author} {\bibfnamefont {S.}~\bibnamefont
  {Reitzenstein}},\ }\href {\doibase 10.1515/nanoph-2018-0222} {\bibfield
  {journal} {\bibinfo  {journal} {Nanophotonics}\ }\textbf {\bibinfo {volume}
  {8}},\ \bibinfo {pages} {655} (\bibinfo {year} {2019})}\BibitemShut {NoStop}%
\bibitem [{\citenamefont {Dorner}\ and\ \citenamefont
  {Zoller}(2002)}]{Dorner2002}%
  \BibitemOpen
  \bibfield  {author} {\bibinfo {author} {\bibfnamefont {U.}~\bibnamefont
  {Dorner}}\ and\ \bibinfo {author} {\bibfnamefont {P.}~\bibnamefont
  {Zoller}},\ }\href {\doibase 10.1103/PhysRevA.66.023816} {\bibfield
  {journal} {\bibinfo  {journal} {Phys. Rev. A}\ }\textbf {\bibinfo {volume}
  {66}},\ \bibinfo {pages} {023816} (\bibinfo {year} {2002})}\BibitemShut
  {NoStop}%
\bibitem [{\citenamefont {Kabuss}\ \emph {et~al.}(2015)\citenamefont {Kabuss},
  \citenamefont {Krimer}, \citenamefont {Rotter}, \citenamefont {Stannigel},
  \citenamefont {Knorr},\ and\ \citenamefont {Carmele}}]{Kabuss2015}%
  \BibitemOpen
  \bibfield  {author} {\bibinfo {author} {\bibfnamefont {J.}~\bibnamefont
  {Kabuss}}, \bibinfo {author} {\bibfnamefont {D.~O.}\ \bibnamefont {Krimer}},
  \bibinfo {author} {\bibfnamefont {S.}~\bibnamefont {Rotter}}, \bibinfo
  {author} {\bibfnamefont {K.}~\bibnamefont {Stannigel}}, \bibinfo {author}
  {\bibfnamefont {A.}~\bibnamefont {Knorr}}, \ and\ \bibinfo {author}
  {\bibfnamefont {A.}~\bibnamefont {Carmele}},\ }\href {\doibase
  10.1103/PhysRevA.92.053801} {\bibfield  {journal} {\bibinfo  {journal} {Phys.
  Rev. A}\ }\textbf {\bibinfo {volume} {92}},\ \bibinfo {pages} {053801}
  (\bibinfo {year} {2015})}\BibitemShut {NoStop}%
\bibitem [{\citenamefont {Carmele}\ \emph
  {et~al.}(2020{\natexlab{b}})\citenamefont {Carmele}, \citenamefont
  {Parkins},\ and\ \citenamefont {Knorr}}]{Carmele2020}%
  \BibitemOpen
  \bibfield  {author} {\bibinfo {author} {\bibfnamefont {A.}~\bibnamefont
  {Carmele}}, \bibinfo {author} {\bibfnamefont {S.}~\bibnamefont {Parkins}}, \
  and\ \bibinfo {author} {\bibfnamefont {A.}~\bibnamefont {Knorr}},\ }\href
  {\doibase 10.1103/PhysRevA.102.033712} {\bibfield  {journal} {\bibinfo
  {journal} {Phys. Rev. A}\ }\textbf {\bibinfo {volume} {102}},\ \bibinfo
  {pages} {033712} (\bibinfo {year} {2020}{\natexlab{b}})}\BibitemShut
  {NoStop}%
\bibitem [{\citenamefont {Grimsmo}(2015)}]{Grimsmo2015}%
  \BibitemOpen
  \bibfield  {author} {\bibinfo {author} {\bibfnamefont {A.~L.}\ \bibnamefont
  {Grimsmo}},\ }\href {\doibase 10.1103/PhysRevLett.115.060402} {\bibfield
  {journal} {\bibinfo  {journal} {Phys. Rev. Lett.}\ }\textbf {\bibinfo
  {volume} {115}},\ \bibinfo {pages} {060402} (\bibinfo {year}
  {2015})}\BibitemShut {NoStop}%
\bibitem [{\citenamefont {Wilson}\ \emph {et~al.}(2003)\citenamefont {Wilson},
  \citenamefont {Bushev}, \citenamefont {Eschner}, \citenamefont
  {Schmidt-Kaler}, \citenamefont {Becher}, \citenamefont {Blatt},\ and\
  \citenamefont {Dorner}}]{Wilson2003}%
  \BibitemOpen
  \bibfield  {author} {\bibinfo {author} {\bibfnamefont {M.~A.}\ \bibnamefont
  {Wilson}}, \bibinfo {author} {\bibfnamefont {P.}~\bibnamefont {Bushev}},
  \bibinfo {author} {\bibfnamefont {J.}~\bibnamefont {Eschner}}, \bibinfo
  {author} {\bibfnamefont {F.}~\bibnamefont {Schmidt-Kaler}}, \bibinfo {author}
  {\bibfnamefont {C.}~\bibnamefont {Becher}}, \bibinfo {author} {\bibfnamefont
  {R.}~\bibnamefont {Blatt}}, \ and\ \bibinfo {author} {\bibfnamefont
  {U.}~\bibnamefont {Dorner}},\ }\href {\doibase 10.1103/PhysRevLett.91.213602}
  {\bibfield  {journal} {\bibinfo  {journal} {Phys. Rev. Lett.}\ }\textbf
  {\bibinfo {volume} {91}},\ \bibinfo {pages} {213602} (\bibinfo {year}
  {2003})}\BibitemShut {NoStop}%
\bibitem [{\citenamefont {H\'etet}\ \emph {et~al.}(2011)\citenamefont
  {H\'etet}, \citenamefont {Slodi\ifmmode~\check{c}\else \v{c}\fi{}ka},
  \citenamefont {Hennrich},\ and\ \citenamefont {Blatt}}]{Hetet2011}%
  \BibitemOpen
  \bibfield  {author} {\bibinfo {author} {\bibfnamefont {G.}~\bibnamefont
  {H\'etet}}, \bibinfo {author} {\bibfnamefont {L.}~\bibnamefont
  {Slodi\ifmmode~\check{c}\else \v{c}\fi{}ka}}, \bibinfo {author}
  {\bibfnamefont {M.}~\bibnamefont {Hennrich}}, \ and\ \bibinfo {author}
  {\bibfnamefont {R.}~\bibnamefont {Blatt}},\ }\href {\doibase
  10.1103/PhysRevLett.107.133002} {\bibfield  {journal} {\bibinfo  {journal}
  {Phys. Rev. Lett.}\ }\textbf {\bibinfo {volume} {107}},\ \bibinfo {pages}
  {133002} (\bibinfo {year} {2011})}\BibitemShut {NoStop}%
\bibitem [{\citenamefont {Hoi}\ \emph {et~al.}(2015)\citenamefont {Hoi},
  \citenamefont {Kockum}, \citenamefont {Tornberg}, \citenamefont
  {Pourkabirian}, \citenamefont {Johansson}, \citenamefont {Delsing},\ and\
  \citenamefont {Wilson}}]{Hoi2015}%
  \BibitemOpen
  \bibfield  {author} {\bibinfo {author} {\bibfnamefont {I.-C.}\ \bibnamefont
  {Hoi}}, \bibinfo {author} {\bibfnamefont {A.~F.}\ \bibnamefont {Kockum}},
  \bibinfo {author} {\bibfnamefont {L.}~\bibnamefont {Tornberg}}, \bibinfo
  {author} {\bibfnamefont {A.}~\bibnamefont {Pourkabirian}}, \bibinfo {author}
  {\bibfnamefont {G.}~\bibnamefont {Johansson}}, \bibinfo {author}
  {\bibfnamefont {P.}~\bibnamefont {Delsing}}, \ and\ \bibinfo {author}
  {\bibfnamefont {C.~M.}\ \bibnamefont {Wilson}},\ }\href {\doibase
  10.1038/nphys3484} {\bibfield  {journal} {\bibinfo  {journal} {Nature
  Physics}\ }\textbf {\bibinfo {volume} {11}},\ \bibinfo {pages} {1045}
  (\bibinfo {year} {2015})}\BibitemShut {NoStop}%
\bibitem [{\citenamefont {Svidzinsky}\ \emph {et~al.}(2018)\citenamefont
  {Svidzinsky}, \citenamefont {Ben-Benjamin}, \citenamefont {Fulling},\ and\
  \citenamefont {Page}}]{Svidzinsky2018}%
  \BibitemOpen
  \bibfield  {author} {\bibinfo {author} {\bibfnamefont {A.~A.}\ \bibnamefont
  {Svidzinsky}}, \bibinfo {author} {\bibfnamefont {J.~S.}\ \bibnamefont
  {Ben-Benjamin}}, \bibinfo {author} {\bibfnamefont {S.~A.}\ \bibnamefont
  {Fulling}}, \ and\ \bibinfo {author} {\bibfnamefont {D.~N.}\ \bibnamefont
  {Page}},\ }\href {\doibase 10.1103/PhysRevLett.121.071301} {\bibfield
  {journal} {\bibinfo  {journal} {Phys. Rev. Lett.}\ }\textbf {\bibinfo
  {volume} {121}},\ \bibinfo {pages} {071301} (\bibinfo {year}
  {2018})}\BibitemShut {NoStop}%
\bibitem [{\citenamefont {Barkemeyer}\ \emph {et~al.}(2019)\citenamefont
  {Barkemeyer}, \citenamefont {Finsterhölzl}, \citenamefont {Knorr},\ and\
  \citenamefont {Carmele}}]{Barkemeyer2019}%
  \BibitemOpen
  \bibfield  {author} {\bibinfo {author} {\bibfnamefont {K.}~\bibnamefont
  {Barkemeyer}}, \bibinfo {author} {\bibfnamefont {R.}~\bibnamefont
  {Finsterhölzl}}, \bibinfo {author} {\bibfnamefont {A.}~\bibnamefont
  {Knorr}}, \ and\ \bibinfo {author} {\bibfnamefont {A.}~\bibnamefont
  {Carmele}},\ }\href {\doibase 10.1002/qute.201900078} {\bibfield  {journal}
  {\bibinfo  {journal} {Advanced Quantum Technologies}\ }\textbf {\bibinfo
  {volume} {3}},\ \bibinfo {pages} {1900078} (\bibinfo {year}
  {2019})}\BibitemShut {NoStop}%
\bibitem [{\citenamefont {Dubin}\ \emph {et~al.}(2007)\citenamefont {Dubin},
  \citenamefont {Rotter}, \citenamefont {Mukherjee}, \citenamefont {Russo},
  \citenamefont {Eschner},\ and\ \citenamefont {Blatt}}]{Dubin2007}%
  \BibitemOpen
  \bibfield  {author} {\bibinfo {author} {\bibfnamefont {F.}~\bibnamefont
  {Dubin}}, \bibinfo {author} {\bibfnamefont {D.}~\bibnamefont {Rotter}},
  \bibinfo {author} {\bibfnamefont {M.}~\bibnamefont {Mukherjee}}, \bibinfo
  {author} {\bibfnamefont {C.}~\bibnamefont {Russo}}, \bibinfo {author}
  {\bibfnamefont {J.}~\bibnamefont {Eschner}}, \ and\ \bibinfo {author}
  {\bibfnamefont {R.}~\bibnamefont {Blatt}},\ }\href {\doibase
  10.1103/PhysRevLett.98.183003} {\bibfield  {journal} {\bibinfo  {journal}
  {Phys. Rev. Lett.}\ }\textbf {\bibinfo {volume} {98}},\ \bibinfo {pages}
  {183003} (\bibinfo {year} {2007})}\BibitemShut {NoStop}%
\bibitem [{\citenamefont {Glaetzle}\ \emph {et~al.}(2010)\citenamefont
  {Glaetzle}, \citenamefont {Hammerer}, \citenamefont {Daley}, \citenamefont
  {Blatt},\ and\ \citenamefont {Zoller}}]{Glaetzle2010}%
  \BibitemOpen
  \bibfield  {author} {\bibinfo {author} {\bibfnamefont {A.}~\bibnamefont
  {Glaetzle}}, \bibinfo {author} {\bibfnamefont {K.}~\bibnamefont {Hammerer}},
  \bibinfo {author} {\bibfnamefont {A.}~\bibnamefont {Daley}}, \bibinfo
  {author} {\bibfnamefont {R.}~\bibnamefont {Blatt}}, \ and\ \bibinfo {author}
  {\bibfnamefont {P.}~\bibnamefont {Zoller}},\ }\href {\doibase
  https://doi.org/10.1016/j.optcom.2009.10.063} {\bibfield  {journal} {\bibinfo
   {journal} {Optics Communications}\ }\textbf {\bibinfo {volume} {283}},\
  \bibinfo {pages} {758 } (\bibinfo {year} {2010})}\BibitemShut {NoStop}%
\bibitem [{\citenamefont {Carmele}\ \emph
  {et~al.}(2013{\natexlab{a}})\citenamefont {Carmele}, \citenamefont {Kabuss},
  \citenamefont {Schulze}, \citenamefont {Reitzenstein},\ and\ \citenamefont
  {Knorr}}]{Carmele2013}%
  \BibitemOpen
  \bibfield  {author} {\bibinfo {author} {\bibfnamefont {A.}~\bibnamefont
  {Carmele}}, \bibinfo {author} {\bibfnamefont {J.}~\bibnamefont {Kabuss}},
  \bibinfo {author} {\bibfnamefont {F.}~\bibnamefont {Schulze}}, \bibinfo
  {author} {\bibfnamefont {S.}~\bibnamefont {Reitzenstein}}, \ and\ \bibinfo
  {author} {\bibfnamefont {A.}~\bibnamefont {Knorr}},\ }\href {\doibase
  10.1103/PhysRevLett.110.013601} {\bibfield  {journal} {\bibinfo  {journal}
  {Phys. Rev. Lett.}\ }\textbf {\bibinfo {volume} {110}},\ \bibinfo {pages}
  {013601} (\bibinfo {year} {2013}{\natexlab{a}})}\BibitemShut {NoStop}%
\bibitem [{\citenamefont {N\'emet}\ \emph {et~al.}(2019)\citenamefont
  {N\'emet}, \citenamefont {Parkins}, \citenamefont {Knorr},\ and\
  \citenamefont {Carmele}}]{Nemet2019}%
  \BibitemOpen
  \bibfield  {author} {\bibinfo {author} {\bibfnamefont {N.}~\bibnamefont
  {N\'emet}}, \bibinfo {author} {\bibfnamefont {S.}~\bibnamefont {Parkins}},
  \bibinfo {author} {\bibfnamefont {A.}~\bibnamefont {Knorr}}, \ and\ \bibinfo
  {author} {\bibfnamefont {A.}~\bibnamefont {Carmele}},\ }\href {\doibase
  10.1103/PhysRevA.99.053809} {\bibfield  {journal} {\bibinfo  {journal} {Phys.
  Rev. A}\ }\textbf {\bibinfo {volume} {99}},\ \bibinfo {pages} {053809}
  (\bibinfo {year} {2019})}\BibitemShut {NoStop}%
\bibitem [{\citenamefont {Pichler}\ \emph {et~al.}(2017)\citenamefont
  {Pichler}, \citenamefont {Choi}, \citenamefont {Zoller},\ and\ \citenamefont
  {Lukin}}]{Pichler2017}%
  \BibitemOpen
  \bibfield  {author} {\bibinfo {author} {\bibfnamefont {H.}~\bibnamefont
  {Pichler}}, \bibinfo {author} {\bibfnamefont {S.}~\bibnamefont {Choi}},
  \bibinfo {author} {\bibfnamefont {P.}~\bibnamefont {Zoller}}, \ and\ \bibinfo
  {author} {\bibfnamefont {M.~D.}\ \bibnamefont {Lukin}},\ } {\bibfield  {journal} {\bibinfo  {journal}
  {Proceedings of the National Academy of Sciences}\ }\textbf {\bibinfo
  {volume} {114}},\ \bibinfo {pages} {11362} (\bibinfo {year}
  {2017})}\BibitemShut {NoStop}%
\bibitem [{\citenamefont {Calaj\'o}\ \emph
  {et~al.}(2019{\natexlab{b}})\citenamefont {Calaj\'o}, \citenamefont {Fang},
  \citenamefont {Baranger},\ and\ \citenamefont {Ciccarello}}]{baranger_2019}%
  \BibitemOpen
  \bibfield  {author} {\bibinfo {author} {\bibfnamefont {G.}~\bibnamefont
  {Calaj\'o}}, \bibinfo {author} {\bibfnamefont {Y.-L.~L.}\ \bibnamefont
  {Fang}}, \bibinfo {author} {\bibfnamefont {H.~U.}\ \bibnamefont {Baranger}},
  \ and\ \bibinfo {author} {\bibfnamefont {F.}~\bibnamefont {Ciccarello}},\
  }\href {\doibase 10.1103/PhysRevLett.122.073601} {\bibfield  {journal}
  {\bibinfo  {journal} {Phys. Rev. Lett.}\ }\textbf {\bibinfo {volume} {122}},\
  \bibinfo {pages} {073601} (\bibinfo {year} {2019}{\natexlab{b}})}\BibitemShut
  {NoStop}%
\bibitem [{\citenamefont {F{\"o}rstner}\ \emph {et~al.}(2003)\citenamefont
  {F{\"o}rstner}, \citenamefont {Weber}, \citenamefont {Danckwerts},\ and\
  \citenamefont {Knorr}}]{Foerstner2003}%
  \BibitemOpen
  \bibfield  {author} {\bibinfo {author} {\bibfnamefont {J.}~\bibnamefont
  {F{\"o}rstner}}, \bibinfo {author} {\bibfnamefont {C.}~\bibnamefont {Weber}},
  \bibinfo {author} {\bibfnamefont {J.}~\bibnamefont {Danckwerts}}, \ and\
  \bibinfo {author} {\bibfnamefont {A.}~\bibnamefont {Knorr}},\ }\href
  {\doibase 10.1103/PhysRevLett.91.127401} {\bibfield  {journal} {\bibinfo
  {journal} {Phys. Rev. Lett.}\ }\textbf {\bibinfo {volume} {91}},\ \bibinfo
  {pages} {127401} (\bibinfo {year} {2003})}\BibitemShut {NoStop}%
\bibitem [{\citenamefont {Förstner}\ \emph {et~al.}(2003)\citenamefont
  {Förstner}, \citenamefont {Weber}, \citenamefont {Danckwerts},\ and\
  \citenamefont {Knorr}}]{Foerstner2003pssb}%
  \BibitemOpen
  \bibfield  {author} {\bibinfo {author} {\bibfnamefont {J.}~\bibnamefont
  {Förstner}}, \bibinfo {author} {\bibfnamefont {C.}~\bibnamefont {Weber}},
  \bibinfo {author} {\bibfnamefont {J.}~\bibnamefont {Danckwerts}}, \ and\
  \bibinfo {author} {\bibfnamefont {A.}~\bibnamefont {Knorr}},\ }\href
  {\doibase 10.1002/pssb.200303155} {\bibfield  {journal} {\bibinfo  {journal}
  {physica status solidi (b)}\ }\textbf {\bibinfo {volume} {238}},\ \bibinfo
  {pages} {419} (\bibinfo {year} {2003})}\BibitemShut {NoStop}%
\bibitem [{\citenamefont {Carmele}\ \emph
  {et~al.}(2013{\natexlab{b}})\citenamefont {Carmele}, \citenamefont {Knorr},\
  and\ \citenamefont {Milde}}]{CarmeleMilde2013}%
  \BibitemOpen
  \bibfield  {author} {\bibinfo {author} {\bibfnamefont {A.}~\bibnamefont
  {Carmele}}, \bibinfo {author} {\bibfnamefont {A.}~\bibnamefont {Knorr}}, \
  and\ \bibinfo {author} {\bibfnamefont {F.}~\bibnamefont {Milde}},\ }\href
  {\doibase 10.1088/1367-2630/15/10/105024} {\bibfield  {journal} {\bibinfo
  {journal} {New Journal of Physics}\ }\textbf {\bibinfo {volume} {15}},\
  \bibinfo {pages} {105024} (\bibinfo {year} {2013}{\natexlab{b}})}\BibitemShut
  {NoStop}%
\end{thebibliography}
\end{document}